%% file: main.tex
\documentclass[submission,copyright,creativecommons]{eptcs}
\providecommand{\event}{NCMA 2023}

\usepackage{amsmath}
\usepackage{amssymb}
\usepackage{amsthm}
\usepackage{csquotes}
\usepackage[inline]{enumitem}
\usepackage{graphicx}
\usepackage{iftex}
\usepackage{isomath}
\usepackage{latexsym}
\usepackage{mathtools}
\usepackage[numbers]{natbib}
\usepackage{stmaryrd}
\usepackage{stackengine}
\usepackage[
    backgroundcolor=orange!5,
    bordercolor=orange,
    linecolor=orange,
    textsize=footnotesize,
]{todonotes}
\usepackage{thmtools}
\usepackage{tikz}
\usepackage{upgreek}
\usepackage{xparse}

\ifpdf
  \usepackage{underscore}         
  \usepackage[T1]{fontenc}        
\else
  \usepackage{breakurl}           
\fi

\usepackage[bb=boondox]{mathalfa}

\setlist[enumerate]{label=(\arabic*)}



\newcommand{\precedes}{\curvearrowright}

\input{math-notation.tex}
\def\textcolon{:}
\catcode`:=13
\def:{\ifmmode\colon\else\textcolon\fi}

\renewcommand*\CTAlg{\alg{C\!T}}

\RenewDocumentCommand\wt{d()}%
 {\mathit{wt}\IfNoValueTF{#1}{}{(#1)}}%

\usetikzlibrary{arrows,positioning,calc,trees,fit}
\usetikzlibrary{tikzmark}
\definecolor{mygray}{gray}{0.6}

\title{Constituency Parsing as an Instance of \\ the M-monoid Parsing Problem}

\author{Richard Mörbitz
  \institute{Faculty of Computer Science \\ Technische Universität Dresden \\ 01062 Dresden}
  \email{richard.moerbitz@tu-dresden.de}
}

\def\titlerunning{Constituency Parsing as an Instance of the M-monoid Parsing Problem}
\def\authorrunning{R. Mörbitz}

\begin{document}
\maketitle
\begin{abstract}
    We consider the constituent parsing problem which states:
    given a final state normalized constituent tree automaton (CTA) and a string,
    compute the set of all constituent trees that are inductively recognized by the CTA
    and yield the string.
    We show that this problem is an instance of the M-monoid parsing problem.
    Moreover, we show that we can employ the generic M-monoid parsing algorithm
    to solve the constituency parsing problem for a meaningful class of CTA.
\end{abstract}

\section{Introduction}

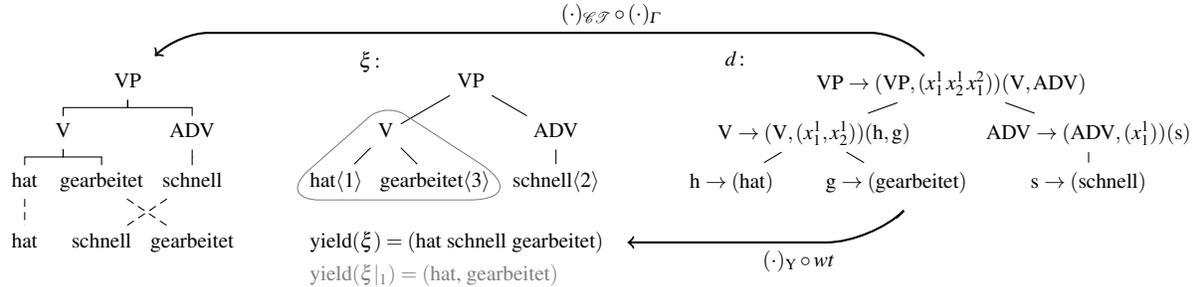
\begin{figure*}[t]
    \centering
    \input{fig1.tex}
    \caption{\label{fig:constituent-trees}%
    Left: constituent tree for the German phrase \enquote{hat schnell gearbeitet}.
    It is discontinuous as the phrase of the left subtree is interleaved
    with a word from the right subtree.
    Center: formalization of this constituent tree in the framework of~\cite{dremoevog22}.
    Right: AST of an RTG and its evaluation in the algebras of our M-monoid parsing problem.}
\end{figure*}

Constituency, sometimes also referred to as phrase structure,
is an important aspect of natural language processing (NLP).
Given a phrase of natural language, the task of constituency parsing
consists in computing a tree-like structure which describes
the syntactic composition of the phrase.
These structures are usually visualized as trees
where the words of the phrase occur as leaves.
Figure~\ref{fig:constituent-trees} (left)
shows such a constituent tree for the German phrase \enquote{hat schnell gearbeitet}.
The ordering of the phrase is indicated below the tree
where dashed lines link the leaves to their corresponding positions in the phrase.
A special phenomenon that may occur in the scope of constituency parsing
are discontinuous constituents.
These span non-contiguous parts of a phrase;
for instance, cf.\@ the constituent labeled $\mathrm{V}$ which spans
the subphrases \enquote{hat} and \enquote{gearbeitet} in our example.
In the usual illustration, discontinuity manifests itself by crossing lines
between the leaves of the tree and the ordering of the phrase.

Usual formal models employed in NLP, such as context-free grammars (CFG)
and finite-state tree automata (FTA),
are not adequate for modeling discontinuous constituents.
This problem has been solved on the grammar side
by exploring more powerful grammar formalisms
such as tree adjoining grammars (TAG; \cite{jossch97}) and
linear context-free rewriting systems (LCFRS; \cite{SekMatFujKas91,kal10}).
On the automaton side,
hybrid tree automata~\cite{dremoevog22} have recently been introduced.
In this context, hybrid trees are usual trees where
labels can be extended by a positive number, called \emph{index},
which indicates their position in the phrase.
(Each index may only occur once per hybrid tree.)
Thus, constituent trees are a particular type of hybrid trees
where a label has an index if and only if it occurs at a leaf position.
Cf.\@ the tree $\xi$ in Fig.~\ref{fig:constituent-trees} (center)
which corresponds to the constituent tree from above.
The previously mentioned discontinuity in its first subtree is resembled by the fact
that the set of indices occurring in this subtree is not contiguous.
Given a constituent tree, we can obtain its phrase
by reading off the labels at the leaves in the order of their indices;
we call this operation \emph{yield}.
Non-contiguous indices lead to phrases with gaps that are formalized using a comma.
For instance, the first subtree of $\xi$ (whose root is labeled by $\mathrm{V}$)
yields the string tuple $(\text{hat}, \text{gearbeitet})$.
In contrast to this formalization of constituent trees,
the usual representation of constituent trees in NLP does not feature indices
and is thus more abstract.

We briefly recall the automaton model of~\cite{dremoevog22}.
In essence, a hybrid tree automaton (HTA) is an FTA where
each transition additionally has an \emph{index constraint}
which describes the acceptable combinations of indices.
Such a constraint may refer to both the indices occurring in the subtrees of the position
where the transition is applied and the index occurring at that position itself.
If unrestricted, these general constraints lead to an overly expressive automaton model.
This is why~\cite{dremoevog22} also introduced constituent tree automata (CTA)
as a restricted form of HTA to
recognize languages of constituent trees.
Here, the index constraints are given by \emph{word tuples}
as they occur in LCFRS.
For instance, the word tuple $(x_1^1 x_2^1 x_1^2)$ states that
the indices of the first subtree form two separate intervals,
i.e., sets of contiguous numbers,
referred to by $x_1^1$ and $x_1^2$,
and the indices of the second subtree ($x_2^1$) lie in between.
Thus, discontinuous constituents can also be modeled.
In essence, a CTA is final state normalized if the constituent trees it recognizes
may only yield contiguous phrases (of course, there may be discontinuity in the subtrees).
Drewes et al.~(2022)~\cite{dremoevog22} showed that the yields of the languages inductively recognized by final state normalized CTA
are equal to the languages generated by LCFRS.
Thus, CTA provide a meaningful framework for specifying constituency analyses.
They did, however, not tackle the following problem which we call
the \emph{constituency parsing problem}:
given a final state normalized CTA $\cA$ and a string $u$,
compute the set of all constituent trees that are inductively recognized
by $\cA$ and yield $u$.
In this paper, we will solve this problem by
showing that it is an instance of the M-monoid parsing problem
to which the generic M-monoid parsing algorithm can be applied,
provided that $\cA$ fulfils a certain condition.

M-monoid parsing~\cite{moevog2019,moevog2021} is
an algebraic framework for weighted parsing.
Its kernel is a \emph{weighted RTG-based language model} (wRTG-LM) $\bar{G}$;
each wRTG-LM consists of a regular tree grammar (RTG) $\cG$,
a $\Gamma$-algebra $(\lalg{L},\phi)$ called \emph{language algebra},
a complete M-monoid $\walg{K}$ called \emph{weight algebra},
and a weight mapping $\wt$ from the set of rules of $\cG$ to the signature of $\walg{K}$.
Moreover, the terminal alphabet of $\cG$ is required to be a subset of $\Gamma$.
The algebraic computations are based on the abstract syntax trees (ASTs) of $\cG$;
these are trees over rules which represent valid derivations.
In the language algebra, each AST can be evaluated to an element of $\lalg{L}$
by first projecting it to a tree over $\Gamma$ and then applying the unique homomorphism
from the $\Gamma$-term algebra to $\lalg{L}$.
In the weight algebra, each AST can be evaluated to an element of $\walg{K}$
by first applying $\wt$ to every rule and then applying the unique homomorphism
from the $\Omega$-term algebra to $\walg{K}$.

The \emph{M-monoid parsing problem} states the following:
given a wRTG-LM $\bar{G}$ and an element $u \in \lalg{L}$ of the language algebra,
compute the sum of the weights (in $\walg{K}$) of all ASTs of $\cG$ which
have the initial nonterminal as the left-hand side of the rule in their root and
evaluate to~$u$ in the language algebra.

Our first contribution is the instantiation of the M-monoid parsing problem
to constituency parsing.
In attempting this instantiation, the constituent trees by \cite{dremoevog22}
turn out to be not suitable for such an algebraic framework.
Instead, we introduce \emph{partitioned constituent trees}
which are inspired by Nederhof and Vogler~(2014)~\cite{nedvog14} (also cf.~\cite{gebnedvog17}).
They are tuples consisting of a (usual) tree,
a strict total order on its leaves,
and a partitioning of its leaves.
Compared to constituent trees, partitioned constituent trees
abstract from particular indices and only preserve information about
the order of the leaves and their groupings (where leaves with consecutive indices
fall into the same component of the partitioning).
This is also closer to the usual notion of constituent trees in NLP.
The yield of partitioned constituent trees is defined analogously to constituent trees:
now, the order of the labels is determined by the total order on the leaves
and commas are placed between labels whose positions belong to different subsets
of the partitioning.

The main part of the instantiation is the following construction.
Given a final state normalized CTA $\cA$, we construct a wRTG-LM $\bar{G}$
such that the M-monoid parsing problem for $\bar{G}$
is equal to the constituency parsing problem for $\cA$.
For the definition of $\bar{G}$, we introduce two algebras:
one for computing partitioned constituent trees
and one for computing their yield.
Both use the same signature $\Gamma$ where
each operator consists of a symbol from some ranked alphabet $\Sigma$ and a word tuple.
The first algebra, called \emph{constituent tree algebra},
operates on partitioned constituent trees over $\Sigma$
by performing top concatenation on their tree components
(using the operator's symbol from $\Sigma$)
and merging their total orders and partitionings using the operator's word tuple.
The second one, called \emph{constituent tree yield algebra},
operates on $\Sigma$-string tuples
by combining them using the operator's word tuple
in the same way as the language generated by an LCFRS is computed.
Both algebras are many-sorted to ensure that the operators and the arguments fit.

Now, given a CTA $\cA$, we define the $\cA$-wRTG-LM $\bar{G}$
as follows.
Its RTG $\cG$ is a syntactical variant of $\cA$,
where the nonterminals, terminals, and the initial nonterminal of $\cG$
are the states of $\cA$, a particular subset of $\Gamma$,
and the final state of $\cA$, respectively.
Each transition $(q_1 \cdots q_k, a, e, q)$ of $\cA$
becomes a rule $q \to (a, e)(q_1, \dots, q_k)$ in $G$.
Moreover, the language algebra of $\bar{G}$ is the constituent tree yield algebra,
the weight algebra is the constituent tree algebra lifted to sets,
and the weight mapping maps each rule to its $\Gamma$-symbol.
This leads to the following M-monoid parsing problem.
Given $\bar{G}$ and $u \in \Sigma^*$,
compute the set of all partitioned constituent trees that
are results of evaluating an AST $d$ of $\cG$ in the constituent tree algebra,
provided that $d$ evaluates to $u$ in the constituent tree yield algebra.
We can prove that this set equals, modulo particular indices,
the set of all constituent trees that are inductively recognized by $\cA$ and yield $u$.
Since adding (resp.\@ removing) these indices is trivial,
the M-monoid parsing problem for $\bar{G}$ and $u$
is equivalent to the constituency parsing problem for $\cA$ and~$u$.
In Figure~\ref{fig:constituent-trees} we indicate the result of this construction
by showing an AST $d$ of some RTG $\cG$ which could be obtained as the result
of the above construction for a given CTA $\cA$ that inductively recognizes $\xi$.
Moreover, we show the evaluation of $d$ in the constituent tree yield algebra as well as
in the constituent tree algebra
(where the partitioned constituent tree is shown via the typical illustration of
constituent trees in NLP)
via the homomorphisms $\lcfrshom$ and $\ctahom$, resp.
(Here, $\projG$ projects rules to their $\Gamma$-symbols.)

Our second contribution concerns the applicability of the generic M-monoid parsing algorithm
\cite{moevog2021} to the M-monoid parsing problem defined above.
We find that the algorithm is in general not applicable,
where the problem lies in monadic cycles:
if the CTA $\cA$ contains transitions of the form $(q_1, a_1, e_1, q_2)$,
$(q_2, a_2, e_2, q_3)$, \dots, $(q_n, a_n, e_n, q_1)$,
then termination of the algorithm is not guaranteed.
Otherwise, if $\cA$ is free of such cycles,
the M-monoid parsing algorithm is applicable to the M-monoid parsing problem
constructed from $\cA$ and thus solves the constituency parsing problem for $\cA$.

This paper is structured as follows.
In Section~\ref{sec:prelim} we fix the basic notions and
repeat some mathematic foundations, especially from the area of algebra.
In Sections~\ref{sec:cta} and~\ref{sec:wrtglm}, we recall the central ideas
of CTA and the M-monoid parsing problem, respectively.
In Section~\ref{sec:main}, we detail the definition of the wRTG-LM $\bar{G}$
we use to model constituency parsing
and we show that the corresponding M-monoid parsing problem is equivalent to the
constituency parsing problem.
Finally, in Section~\ref{sec:algorithm},
we discuss the applicability of the M-monoid parsing algorithm.

\section{Preliminaries}
\label{sec:prelim}

\paragraph{Mathematical notions.}
The set of natural numbers (including $0$) is denoted by $\Nat$
and we let $\Nat_+ = \Nat\setminus\{0\}$.
For every $k,\ell \in \Nat$,
we let $[k,\ell]$ denote the interval $\{i \in \Nat \mid k \le i \le \ell\}$ and
we abbreviate $[1,\ell]$ by $[\ell]$.
The set of all nonempty intervals of $\Nat_+$ is denoted by $\ints$.
For every $I,I'\in\ints$, the expression $I<I'$ holds if $\max I <\min I'$
and $I\precedes I'$ holds if $\max I+1=\min I'$.
Thus $I \precedes I'$ implies $I < I'$.
For each set $A$, we let $\PowerSet{A}$ denote the power set of $A$.
We extend a mapping $f: A \to B$ in the canonical way to a mapping
$f: \PowerSet{A} \to \PowerSet{B}$.
A \emph{family} $(a_i \mid i \in I)$ is a mapping $f: I \to A$
with $f(i) = a_i$ for each $i \in I$.
Let $A$, $B$, and $C$ be sets.
The composition of two mappings $f\colon A \to B$ and $g\colon B \to C$
is denoted by $g \circ f$.
Whenever we deal with a partitioning $(A_1, \ldots, A_n)$ of a set $A$,
we require $A_i$ to be non-empty (for each $i \in [n]$).
An \emph{alphabet} is a finite and non-empty set.

\paragraph{Strings and tuples.}
Let $A$ be a set and $k\in\Nat$.
We let $A^k$ denote the set of all strings $w=a_1\cdots a_k$ of length~$k$,
where $a_1, \ldots, a_k \in A$, and we let $A^*=\bigcup_{k\in\Nat}A^k$.
The empty string ($k=0$) is denoted by $\varepsilon$.
We denote substrings of a string $w=a_1\cdots a_k$ in $A^*$
as follows: for every $i \in [k]$ and $j \in [k-i+1]$,
we let $w[i;j] = a_i \cdots a_{i+j-1}$, and $w[i]$ abbreviates $w[i;1]$.
Let $\ell \in \Nat$.
The \emph{concatenation} of two strings $v = a_1 \cdots a_k$ in $A^k$ and $w = b_1 \cdots b_\ell$ in $A^\ell$,
denoted by $v \cdot w$, is the string $a_1 \cdots a_k b_1 \cdots b_\ell$ in $A^{k+\ell}$;
we drop $\cdot$ if it is clear from the context.
Moreover, we lift concatenation to sets of strings in the obvious way.

We let $\Tup[k](A)$ denote the $k$-fold Cartesian product of $A$;
its elements are called \emph{$k$-tuples over $A$}.
Moreover, we let $\Tup(A) = \bigcup_{k \in \Nat} \Tup[k](A)$.
In the obvious way, we transfer the notion of substrings from strings to tuples.

\paragraph{Sorted sets, trees, and regular tree grammars.}
Let $S$ be a set; its elements are usually called \emph{sorts}.
An \emph{$S$-sorted set} is a pair $(A, \sort)$ where $A$ is a set and
$\sort: A \to S$ is a mapping.
For each $s \in S$, we let $A^{(s)} = \{ a \in A \mid \sort(a) = s \}$.
We call an $S$-sorted set \emph{single-sorted} if $|S| = 1$;
thus, each (usual) set can be viewed as a single-sorted set.
A \emph{ranked set} is an $\Nat$-sorted set; its sort mapping is usually denoted by $\rk$.
In examples, we will show the rank of a symbol as a superscript in parentheses,
e.g., $a^{(k)}$ if $\rk(a) = k$.
An $S$-sorted (resp\@. ranked) alphabet is an $S$-sorted (resp.\@ ranked) set
which is an alphabet. 

An $(S^* \times S)$-sorted set $\Gamma$ is called \emph{$S$-signature}.
Whenever we write $\gamma \in \Gamma^{(s_1 \cdots s_k, s)}$ we assume that
$k \in \Nat$ and $s, s_1, \ldots, s_k \in S$ are universally quantified
if not specified otherwise.
Now let $H$ be an $S$-sorted set.
The set of \emph{$S$-sorted trees over $\Gamma$ and $H$}, denoted by $\T[\Gamma](H)$,
is the smallest $S$-sorted set $T$ such that,
        for each $s \in S$, we have $H^{(s)} \subseteq T^{(s)}$ and,
        for every $\gamma \in \Gamma^{(s_1 \cdots s_k, s)}$ and
        $t_1 \in T^{(s_1)}, \ldots, t_k \in T^{(s_k)}$,
        we have $\gamma(t_1, \ldots, t_k) \in T^{(s)}$.
We abbreviate $\T[\Gamma](\emptyset)$ by $\T[\Gamma]$.
Since we can view each $(S^* \times S)$-sorted set as a ranked set by,
for every $\gamma \in \Gamma^{(s_1 \cdots s_k, s)}$, letting $\rk(\gamma) = k$,
the above definition also covers the usual trees over ranked alphabets.

The \emph{set of positions} of a tree is defined by the mapping
$\pos\colon \T[\Gamma](H) \to \PowerSet{(\Nat_+)^*}$ as usual.
Let $t \in \T[\Gamma](H)$ and $w \in \pos(t)$.
The set of \emph{leaves} of $t$, the \emph{label of $t$ at  $w$},
and the \emph{subtree of $t$ at $w$}
are also defined as usual,
and are denoted by $\leaves(t)$, $t(w)$, and $t|_w$, respectively.

An \emph{$S$-sorted regular tree grammar} (RTG; \cite{Brainerd1969})
is a tuple $\cG=(N,\Gamma,A_0,R)$
where $N$ is an $S$-sorted alphabet (\emph{nonterminals}),
$\Gamma$ is an $(S^* \times S)$-sorted alphabet (\emph{terminals})
with $N \cap \Gamma = \emptyset$,
$A_0 \in N$ (\emph{initial nonterminal}),
and $R$ is a finite set of \emph{rules} where each rule $r$ has the form
$A \to \gamma(A_1,\ldots,A_k)$
with $k \in \Nat$, $\gamma \in \Gamma^{(s_1 \cdots s_k,s)}$,
and $A \in N^{(s)}, A_1 \in N^{(s_1)}, \ldots, A_k \in N^{(s_k)}$.
(Thus, we only consider RTGs in normal form.)
We call $A$ the \emph{left-hand side} of $r$; it is denoted by $\lhs(r)$.

We view $R$ as an $(N^* \times N)$-sorted set
where each rule $A \to \gamma(A_1,\ldots,A_k)$ has sort $(A_1 \cdots A_k, A)$.
Thus, for every $d \in \T[R]$ and $w \in \pos(d)$, the following holds:
if $d(w)$ is $A \to \gamma(A_1,\ldots,A_k)$,
then, for each $i \in [k]$, we have $\lhs(d(w \cdot i)) = A_i$.
We call $\T[R]$ the set of \emph{abstract syntax trees} (short: ASTs) of $\cG$.
We define the mapping $\projG: \T[R] \to \T[\Gamma]$ such that
$\projG(d)$ is obtained from $d$
by replacing each $A \to \gamma(A_1,\ldots,A_k)$ by $\gamma$.
The \emph{tree language generated by $\cG$} is the set $\Lang(\cG) = \projG(\T[R])$.

\paragraph{$S$-sorted $\Gamma$-algebras.}
Let $S$ be a set and $\Gamma$ be an $S$-signature.
An \emph{$S$-sorted $\Gamma$-algebra} (short: algebra) is a pair $(\alg A,\phi)$ where
$\alg A$ is an $S$-sorted set (\emph{carrier set})
and $\phi$ is a mapping
which maps each $\gamma \in \Gamma^{(s_1 \cdots s_k, s)}$ to a mapping
$\phi(\gamma): \alg A^{(s_1)} \times \cdots \times \alg A^{(s_k)} \to \alg A^{(s)}$. 
We will sometimes identify $\phi(\gamma)$ and $\gamma$ (as it is usual).

The \emph{$S$-sorted $\Gamma$-term algebra} is the $S$-sorted $\Gamma$-algebra
$(\T[\Gamma],\phi_\Gamma)$ where,
for every $\gamma \in \Gamma^{(s_1 \cdots s_k, s)}$ and
$t_1\in\T[\Gamma]^{(s_1)}, \ldots, t_k \in \T[\Gamma]^{(s_k)}$,
we let
$\phi_\Gamma(\gamma)(t_1,\ldots,t_k) = \gamma(t_1,\ldots,t_k)$.
For each $\Gamma$-algebra $(\alg A,\phi)$ there is a unique homomorphism,
denoted by $\proj{\alg A}$, from the $\Gamma$-term algebra to $(\alg A, \phi)$
\cite{wec92}.
We write its application to an argument $t \in \T[\Gamma]$ as $\proj{\alg A}(t)$.
Intuitively, $\proj{\alg A}$ evaluates a tree $t$ in $(\alg A, \phi)$,
in the same way as arithmetic expressions are evaluated to numbers.
For instance, the expression $3 + 2 \cdot (4 + 5)$
is evaluated to $21$ in the $\{+,\cdot\}$-algebra $(\Nat,+,\cdot)$.
Often we abbreviate an algebra $(\alg A,\phi)$ by $\alg A$.
For every $a \in \alg A$ we let
$\factors(a) = \{ b \in \alg A \mid {b \ltfac}^* a \}$
where, for every $a, b \in \alg A$, $b \ltfac a$
if there is a $\gamma \in \Gamma$ such that $b$ occurs in some tuple $(b_1, \ldots, b_k)$
with $\phi(\gamma)(b_1, \ldots, b_k) = a$.
That is, $\factors(a)$ is the set of all values that occur in a term which evaluates
to $a$.
We call $(\alg A, \phi)$ \emph{finitely decomposable}
if $\factors(a)$ is finite for every $a \in \alg A$.

\paragraph{Word tuples.}
Let $k \in \Nat$ and $\kappa = (\ell_1,\ldots,\ell_k)$ in $\Tup[k](\Nat_+)$.
We let $\Var = \{x_i^j \mid i \in [k], j \in [\ell_i]\}$
and call each element $x_i^j$  of $\Var$ a \emph{variable}.
Moreover, let $n \in \Nat_+$ and $\Delta$ be an alphabet.
Then we denote by $\W(\Delta)$ the set of all tuples $e = (s_1,\ldots,s_n)$ such that
\begin{enumerate*}
    \item for each $i \in [n]$,
        the component $s_i$ is a string over $\Delta$ and $\Var$,
    \item each variable in $\Var$ occurs exactly once in $e$, and
    \item for all $x_i^{j_1},x_i^{j_2} \in \Var$ with $j_1 < j_2$,
        the variable $x_i^{j_1}$ occurs left of $x_i^{j_2}$ in $e$.
\end{enumerate*}
Each element of $\W(\Delta)$ is a \emph{monotone $(n,\kappa)$-word tuple}.%
\footnote{Monotonicity is expressed by condition (3); in this paper, we do not deal with non-monotone word tuples.}
We let $\WE(\Delta) = \bigcup_{n \in \Nat_+, k \in \Nat, \kappa \in \Tup[k](\Nat_+)} \W(\Delta)$
and we drop `($\emptyset$)' for empty $\Delta$. 

Let $e = (s_1,\ldots,s_n)$ be in $\W(\Delta)$.
The \emph{word function induced by $e$} is the mapping
\[
    \sem(e): \Tup[\ell_1](\Delta^*) \times \dots \times \Tup[\ell_k](\Delta^*) \to \Tup[n](\Delta^*)
\]
which is defined, for every
$(w_1^1,\ldots,w_1^{\ell_1}) \in \Tup[\ell_1](\Delta^*)$,
\ldots,
$(w_k^1,\ldots,w_k^{\ell_k}) \in \Tup[\ell_k](\Delta^*)$,
by
\[
    \sem(e)((w_1^1,\ldots,w_1^{\ell_1}), \ldots, (w_k^1,\ldots,w_k^{\ell_k})) = (v_1,\ldots,v_n)
\]
where each $v_m$ ($m\in[n]$)
is obtained from $s_m$ by replacing every occurrence of a variable $x_i^j$ by $w_i^j$.
For instance, let $\Delta = \{a,b,c,d\}$.
The word tuple $e = (bx_2^1x_1^1ax_1^2,acx_2^2x_1^3a)$ in
$\W[2][(3,2)](\Delta)$  induces the  word function
\(\sem(e)\colon \Tup[3](\Delta^*) \times \Tup[2](\Delta^*) \to \Tup[2](\Delta^*)\) with
\[
    \sem(e)((w_1^1,w_1^2,w_1^3),(w_2^1,w_2^2)) = (bw_2^1w_1^1aw_1^2,acw_2^2w_1^3a).
\]

We view $\WE(\Delta)$ as a $(\Nat_+^* \times \Nat_+)$-sorted set in the obvious way
(i.e., $e \in \W(\Delta)$ has sort $(\kappa,n)$)
and we denote the unique homomorphism
from the $\Nat_+$-sorted $\WE(\Delta)$-term algebra
to the $\Nat_+$-sorted algebra $(\Tup(\Delta^*), \sem)$
also by $\sem$.
Intuitively, it evaluates trees over word tuples to
elements of $\Tup(\Delta^*)$ by applying in a bottom-up way
the word functions induced by their word tuples.

\paragraph{Monoids.}
A \emph{monoid} is an algebra $(\walg K,\oplus,\BZero)$
such that $\oplus$ is a binary, associative operation on $\walg K$ and
$\BZero \oplus \welem k = \welem k = \welem k \oplus \BZero$
for each $\welem k \in \walg K$.
The monoid is \emph{commutative} if $\oplus$ is commutative and
it is \emph{idempotent} if $\welem k \oplus \welem k = \welem k$.
It is \emph{complete} if, for each countable set $I$,
there is an operation $\sum^{\oplus}_I$ which maps each family $(\welem k_i \mid i \in I)$
to an element of $\walg K$,
coincides with $\oplus$ when $I$ is finite,
and otherwise satisfies axioms which guarantee commutativity and associativity
\cite[p.~124]{Eilenberg1974}.
We abbreviate $\sum^\oplus_I(\welem k_i \mid i \in I)$ by
$\sum^\oplus_{i \in I}\welem k_i$.
A complete monoid is \emph{d-complete}~\cite{Karner1992} if, for every
$\welem k \in \walg K$ and family $(\welem k_i \mid i \in \Nat)$ of elements of~$\walg K$,
the following holds:
if there is an $n_0 \in \Nat$ such that for every $n \in \Nat$ with $n \ge n_0$,
$\infsum_{\substack{i \in \Nat: i \le n}} \welem k_i = \welem k$,
then $\infsum_{i \in \Nat} \welem k_i = \welem k$.
A complete monoid is \emph{completely idempotent}
if for every $\welem k \in \walg K$ and countable set~$I$
it holds that $\infsum_{i \in I} \welem k = \welem k$.
An easy proof shows that if $\walg{K}$ is completely idempotent, it is also d-complete.

\paragraph{M-monoids.}
A \emph{multioperator monoid} (M-monoid; \cite{Kuich1999}) is an algebra
$(\walg K, \oplus,\BZero,\Omega,\phi)$ where
$(\walg K, \oplus,\BZero)$ is a commutative monoid (\emph{additive monoid}),
$\Omega$ is a ranked set, and $(\walg K,\phi)$ is an $\Omega$-algebra.
An M-monoid inherits the properties of its monoid (e.g., being complete). 
We denote a complete M-monoid by $(\walg K, \oplus, \BZero, \Omega, \phi, \infsum)$.
An M-monoid is \emph{distributive} if,
for every $\omega \in \Omega^{(m)}$, $i \in [m]$, and
$\welem{k}, \welem{k}_1, \ldots, \welem{k}_m \in \walg{K}$,
\[
    \omega(\welem{k}_1, \dots, \welem{k}_{i-1}, \welem{k}_i \oplus \welem{k}, \welem{k}_{i+1}, \dots, \welem{k}_m)
    = \omega(\welem{k}_1, \dots, \welem{k}_{i-1}, \welem{k}_i, \welem{k}_{i+1}, \dots, \welem{k}_m)
    \oplus \omega(\welem{k}_1, \dots, \welem{k}_{i-1}, \welem{k}, \welem{k}_{i+1}, \dots, \welem{k}_m).
\]
If $\walg K$ is complete, then we only call it distributive
if the above equation
also holds for each countable set of summands.
We sometimes refer to an M-monoid only by its carrier set.

\begin{example}\label{ex:mmonoid}
    Let $S$ be a set, $\Omega$ be an $S$-signature,
    and $(\alg A, \phi)$ be an $S$-sorted set.
    We will now define an M-monoid which lifts the computations
    from $(\alg A, \phi)$ to sets of elements of $\alg A$.
    Its carrier set will be
    $B = \bigcup_{s \in S} \PowerSet{\alg A^{(s)}} \cup \{ \bot \}$
    where $\bot$ is a new element.
    Thus, $B$ contains all single-sorted subsets of $\alg A$
    and an element $\bot$ which will be used
    whenever an operation is applied to arguments which do not match its sort.
    Formally, we define the M-monoid $(B, \sortedcup, \emptyset, \Omega, \psi)$
    where, for every $B_1, B_2 \in B$,
    \[
        B_1 \sortedcup B_2 = \begin{cases}
            B_1 \cup B_2 &\text{if there exists $s \in S$ such that $B_1, B_2 \subseteq \alg A^{(s)}$} \\
            \bot &\text{otherwise}
        \end{cases}
    \]
    and, for every $\gamma \in \Gamma^{(s_1 \cdots s_k,s)}$ and $B_1, \dots, B_k \in B$,
    \[
        \psi(\gamma)(B_1, \dots, B_k) = \begin{cases}
            \phi(\gamma)(B_1, \dots, B_k) &\text{if $B_1 \subseteq \alg A^{(s_1)}, \dots, B_k \subseteq \alg A^{(s_k)}$} \\
            \bot &\text{otherwise.}
        \end{cases}
    \]
    We consider $\infsum[\sortedcup]$ which is defined for every index set $I$
    as $\bigcup_I$.
    It is easy to see that $B$ together with $\infsum[\sortedcup]$ is complete
    and distributive.
    Moreover, since the monoid $(B, \sortedcup, \emptyset, \infsumop[\sortedcup])$
    is completely idempotent,
    we obtain that $(B, \sortedcup, \emptyset, \Omega, \psi, \infsum[\sortedcup])$ is d-complete.
\end{example}

\section{Constituent tree automata}
\label{sec:cta}

Hybrid trees and, as a special case thereof, constituent trees
are certain trees over potentially indexed symbols where, intuitively,
an indexed symbol is a symbol equipped with a positive number.
Formally, let $\Sigma$ be an alphabet.
The \emph{set of indexed $\Sigma$-symbols}, denoted by $\ind{\Sigma}{\Nat_+}$,
is the ranked set defined by
$\ind{\Sigma}{\Nat_+}^{(k)} \allowbreak = \allowbreak \{\ind{a}{n} \mid a \in \Sigma^{(k)}, n \in \Nat_+\}$ for each $k \in \Nat$.
An element $\ind{a}{n}$ is called \emph{indexed symbol} and
$n$ is the \emph{index of $\ind{a}{n}$}.
We write $\projS(\ind{a}{n})$ for $a$ and $\projN(\ind{a}{n})$ for $n$.

Here we only define constituent trees;
for a general definition of hybrid trees, cf.~\cite{dremoevog22}.
A \emph{constituent tree} is a tree $\xi \in \T[\Sigma](\ind\Sigma{\Nat_+})$
such that, for every $w, w' \in \leaves(\xi)$, we have that
$\projN(\xi(w)) = \projN(\xi(w'))$ implies $w = w'$.
In words,
a symbol is indexed if and only if it occurs at a leaf and no index occurs twice.
We let $\projS(\xi)$ denote the tree in $\T$ obtained from $\xi$
by removing all indices.
The set of all constituent trees over $\Sigma$ is denoted by $\CT$.

We extract the linear phrase from a constituent tree $\xi$
using the mapping $\yield: \CT \to \Tup(\Sigma^*)$
which we define as follows.
We order the set of indexed symbols occurring in $\xi$ into a sequence
according to their indices,
then we drop each comma between neighbored symbols with consecutive indices,
and finally we drop the indices.
Thus, the tuple $\yield(\xi)$ has one more component than
the number of gaps in the set of indices occurring in~$\xi$.
For instance, consider the constituent tree $\xi$ in Figure~\ref{fig:constituent-trees}.
The ordering of its set of indexed symbols is
$(\ind{\mathrm{hat}}{1}, \ind{\mathrm{schnell}}{2}, \ind{\mathrm{gearbeitet}}{3})$
and all commas are dropped as there are no gaps between indices.

A \emph{constituent tree automaton} (short: CTA) is
a tuple $\cA = (Q, \Sigma, \delta, q_f)$ where
\begin{itemize}
    \item $Q$ is a ranked alphabet with $Q^{(0)} = \emptyset$ (\emph{states}),
    \item $\Sigma$ is a ranked alphabet,
    \item $\delta$ is a finite set of \emph{transitions},
        each of which having either
            form $(\varepsilon, a, q)$ where $a \in \Sigma^{(0)}$ and $q \in Q^{(1)}$ or
            form $(q_1 \cdots q_k, a, e, q)$
                where $k \in \Nat_+$,
                $e \in \W[n][(\ell_1, \ldots, \ell_k)]$,
                $q_1 \in Q^{(\ell_1)}, \ldots, q_k \in Q^{(\ell_k)},\allowbreak q \in Q^{(n)}$,
                and 
                $a \in \Sigma^{(k)}$;
                and
    \item $q_f \in Q$ (\emph{final state}).
\end{itemize}
We call $\cA$ \emph{final state normalized} if $q_f \in Q^{(1)}$.

We note that this definition of CTA simplifies the definition by~\cite{dremoevog22}
in three regards.
First, we opted to define CTA directly and not as a special case of HTA.
Second, their nullary transitions contain an additional object,
the universal index constraint $\ICAll_{0,1}$,
which we have dropped for the sake of clarity.
Third, to achieve coherence with RTGs, our CTA has only a single final state $q_f$.
This is not a restriction, since each CTA of~\cite{dremoevog22} with a set of final states
can be transformed into an equivalent CTA with a single final state
using a standard construction from automata theory
(cf., e.g., \cite[L.\,4.8]{DroPecVog2005}).

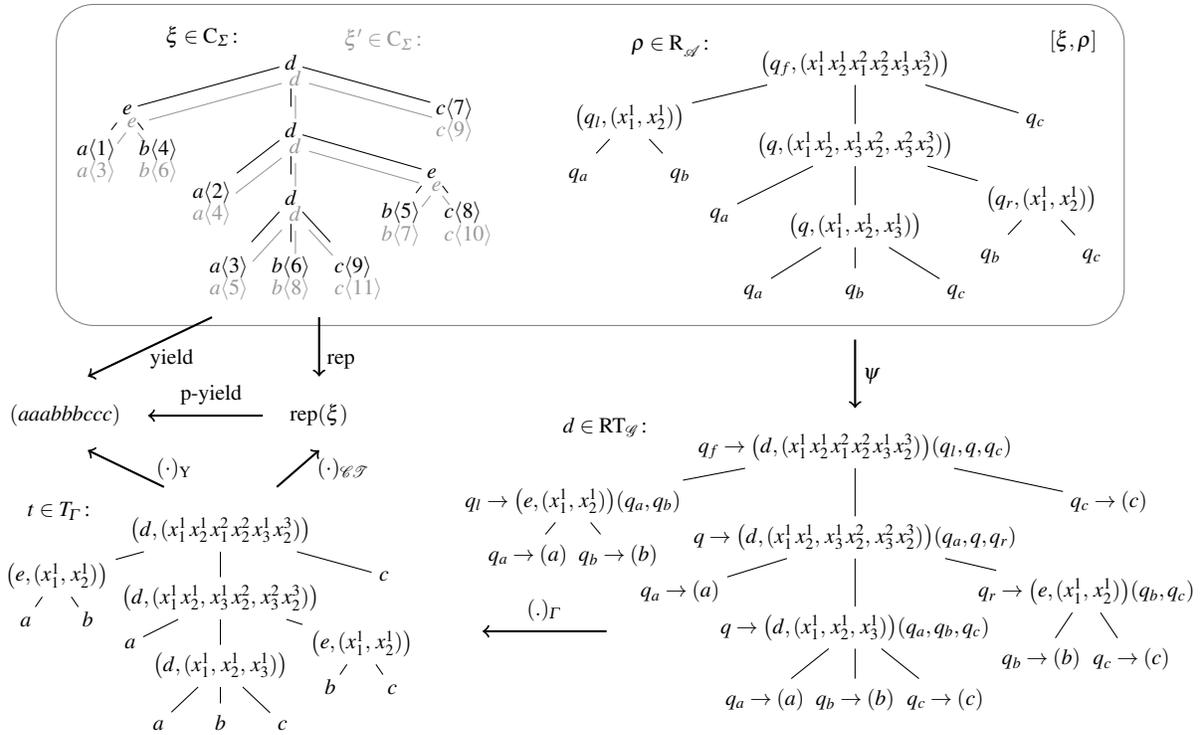
\begin{figure*}[t]
    \centering
    \input{fig2-tighter.tex}
    \caption{\label{fig:main}%
    Top: constituent tree $\xi$ and run $\rho$ of the CTA $\cA$ from Example~\ref{ex:cta}
    such that $(\xi,\rho) \in \LR$.
    We also show a constituent tree $\xi'$ with $(\xi',\rho) \eqrel (\xi,\rho)$ in gray.
    Bottom: AST $d$ of the $\cA$-RTG $\cG$ which is the image of $\eqcl{\xi,\rho}$
    under the bijection $\bij$.
    The mappings introduced in this paper commute;
    in particular, $\yield(\xi) = \lcfrshom(\projG(\bij(\eqcl{\xi,\rho})))$.}
\end{figure*}

\begin{example}\label{ex:cta}
    Let $\cA = (Q, \Sigma, \delta, q_f)$ be a CTA where
    the states are
    $Q = \{ q^{(3)}, q_l^{(2)}, q_r^{(2)}, q_a^{(1)}, q_b^{(1)}, q_c^{(1)}, q_f^{(1)} \}$,
    the terminal alphabet is
    $\Sigma = \{ a^{(0)}, b^{(0)}, c^{(0)}, d^{(3)}, e^{(2)} \}$, and
    $\delta$ consists of the following transitions:
    \[
        \begin{gathered}
            (q_l q q_c, d, (x_1^1 x_2^1 x_1^2 x_2^2 x_3^1 x_2^3), q_f) \qquad
            (q_a q q_r, d, (x_1^1 x_2^1 x_3^1 x_2^2 x_3^2 x_2^3), q_f) \\
            (q_l q q_c, d, (x_1^1 x_2^1, x_1^2 x_2^2, x_3^1 x_2^3), q) \qquad
            (q_a q q_r, d, (x_1^1 x_2^1, x_3^1 x_2^2, x_3^2 x_2^3), q) \\
            (q_a q_b q_c, d, (x_1^1, x_2^1, x_3^1), q) \qquad
            (q_a q_b, e, (x_1^1, x_2^1), q_l) \qquad
            (q_b q_c, e, (x_1^1, x_2^1), q_r) \\
            (\varepsilon, a, q_a) \qquad
            (\varepsilon, b, q_b) \qquad
            (\varepsilon, c, q_c).
        \end{gathered}
    \]
    We note that $\cA$ is final state normalized.
    We will use $\cA$ to illustrate the semantics of CTA which we define next.
\end{example}

While there are two semantics of CTA in~\cite{dremoevog22},
we are only interested in one of them,
called the \emph{hybrid tree language inductively recognized} by CTA\@.
In this paper, we refer to it simply as \emph{language inductively recognized by $\cA$}
and define it in the following.

Let $k, \ell_1, \ldots, \ell_k \in \Nat_+$.
We let $\kappa = (\ell_1, \ldots, \ell_k)$.
A \emph{$\kappa$-assignment} is a mapping $\ass: \Var \to \ints$ such that,
for every $x, x' \in \Var$ with $x \not= x'$,
it holds that $\ass(x) \cap \ass(x') = \emptyset$.
Now let $n \in \Nat_+$ and $e \in \W$.
We say that \emph{$\ass$ models $e$}, denoted by $\ass \models e$,
if the expression $e'$ holds where $e'$ is obtained from $e$ by
\begin{enumerate*}
    \item writing $\precedes$ between each occurrence of two consecutive variables,
    \item replacing each comma by $<$,
        and
    \item replacing each variable $x$ by $\ass(x)$.
\end{enumerate*}
As an example, consider the word tuple
$e =(x^1_1 \, x^1_2, \, x^1_3 \, x^2_2, \, x^2_3 \,  x^3_2)$
which occurs at position $2$ of $\rho$ in Figure~\ref{fig:main}.
We define the $(1,3,2)$-assignment $\ass$ with $\ass(x_1^1) = \{2\}$,
$\ass(x_2^1) = \{3\}$,
$\ass(x_3^1) = \{5\}$,
$\ass(x_2^2) = \{6\}$,
$\ass(x_3^2) = \{8\}$,
$\ass(x_2^3) = \{9\}$,
where the indices are taken from the constituent tree $\xi$ in Figure~\ref{fig:main}.
We obtain the expression
$e' =(\{2\} \precedes \{3\} < \{5\} \precedes \{6\} < \{8\} \precedes \{9\})$
which is valid and hence $\ass \models e$.

Let $\cA = (Q,\Sigma,\delta,q_f)$ be a CTA.
A \emph{run of $\cA$} is a tree $\rho \in \T[Q \times \WE](Q)$
where, for $(q, e) \in Q \times \WE$, we let $\rk(q, e) = k$
if $e \in \W[n][(\ell_1,\ldots,\ell_k)]$.
We let $\runs{\cA}$ denote the set of runs of $\cA$.
We define $\indset \subseteq \CT \times \runs{\cA} \times \Tup(\ints)$
to be the smallest set $T$ that satisfies the following:

\begin{itemize}
    \item For every $(\varepsilon, a, q) \in \delta$ and $i \in \Nat_+$
        it holds that
        \(
            \big(\ind{a}{i}, q, \{i\}\big) \in T.
        \)
    \item 
        For every $(q_1 \cdots q_k, a, e, q) \in \delta$ and
        $(\xi_1, \rho_1, J_1), \allowbreak \ldots, \allowbreak (\xi_k, \rho_k, J_k) \in T$
        where $q_i$ is the state at $\rho_i(\varepsilon)$ (for $i \in [k]$),
        we let $\kappa$ denote
        $(\rk(q_1), \ldots, \rk(q_k))$ and
        consider the mapping
        \(
          \ass\colon \Var \to \ints
        \)
        defined, for every $i \in [k]$ and $j \in [\rk(q_i)]$,
        by $\ass(x^j_i) = J_i[j]$.
        (The fact that $J_i[j]$ is indeed an interval can easily be verified by induction.)
        Now, if $\ass$ is a $\kappa$-assignment
        (i.e., its image consists of pairwise disjoint sets)
        and ${\ass \models e}$, then
        $\big(a(\xi_1,\ldots,\xi_k), \rho, (U_1, \ldots, U_{\rk(q)})\big) \in T$
        where we let $\rho = (q,e)(\rho_1, \ldots, \rho_k)$
        and, for each $m \in [\rk(q)]$,
        $U_m = \bigcup_{\text{$i,j$: $x_i^j$ occurs in the $m$-th component of $e$}} \ass(x_i^j)$.
\end{itemize}
We define the following projection of $\indset$
(where $\LR[{}]$ stands for ``constituent (trees and) runs''):
\[
    \LR = \{ (\xi, \rho) \mid (\exists J \in \Tup(\ints))\ldotp (\xi, \rho, J) \in \indset \}.
\]
The language \emph{inductively recognized by $\cA$},
denoted by $\Lind(\cA)$, is the set
\[
    \Lind(\cA) = \{\xi \mid (\xi,\rho) \in \LR, \rho(\varepsilon) \ \text{has state} \ q_f\}.
\]

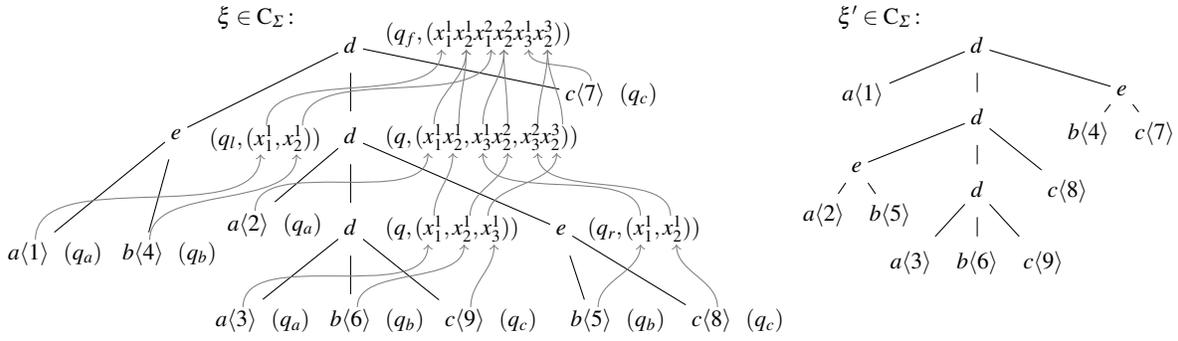
\begin{figure*}[t]
    \centering
    \input{fig-inductive.tex}
    \caption{\label{fig:inductive}%
    Left: constituent tree $\xi$ and run $\rho$ of the CTA $\cA$ from Example~\ref{ex:cta}
    such that $(\xi,\rho) \in \LR$ where the states and word tuples of $\rho$
    have been written next to the positions of $\xi$.
    Arrows indicate the family of assignments which witnesses $(\xi,\rho) \in \LR$
    where, at each non-leaf position of $\rho$, a variable is assigned
    the set of all indices whose arrows reach it.
    Right: another constituent tree $\xi' \in \Lind(\cA)$
    such that $\yield(\xi) = \yield(\xi')$.}
\end{figure*}

\begin{example}
    Recall the CTA $\cA$ of Example~\ref{ex:cta}.
    The top left of Figure~\ref{fig:main} shows a constituent tree $\xi$
    and a run $\rho$ of $\cA$ such that $(\xi, \rho) \in \LR$.
    In order to show that $(\xi, \rho) \in \LR$ indeed holds,
    in Figure~\ref{fig:inductive} (left),
    we illustrate the assignments used at each position of $\xi$
    in the inductive definition of $\indset$.
    For this, we use arrows starting at the indices in the leaves of $\xi$.
    At every non-leaf position $w$ of $\rho$,
    we show the $\kappa$-assignment $\ass$
    which witnesses the existence of $J \in \Tup(\ints)$
    such that $(\xi|_w, \rho|_w, J) \in \indset$ as follows:
    for each variable $x_i^j$ in the word tuple at $\rho(w)$,
    it holds that $\ass(x_i^j)$ consists of all indices whose arrows reach $x_i^j$.
    In the way these arrows pass through the word tuples at subtrees of $\rho|_w$,
    it is shown that $\ass$ is consistent with the assignments in the subtrees.
    This stresses the inductive nature of $\LR$.

    The constituent tree $\xi$ exemplifies the form of each constituent tree
    inductively recognized by $\cA$.
    The backbone is a monadic chain where each position is labeled with $d$.
    The bottom of the chain has three leaf children, labeled by $a$, $b$, and $c$.
    Each inner position of the chain has three children as well,
    the second of which continues the chain.
    Moreover, the symbols $a$, $b$, and $c$ are distributed as leaves
    among the first and third child,
    where $e$ serves as an intermediate node under the child receiving two symbols
    (cf.\@ positions $\varepsilon$ and $2$ of $\xi$).
    The indices are placed such that, for each of $a$, $b$, and $c$,
    the indices occurring with this symbol form an interval
    where $a$ has the lowest and $c$ has the highest interval.
    Thus, $\yield(\Lind(\cA)) = \{ a^nb^nc^n \mid n \in \Nat_+ \}$
    which is not context-free.
    As there are two patterns for inner positions of the backbone,
    $\cA$ may recognize several constituent trees with the same yield
    (an example is given in the right of Figure~\ref{fig:inductive}).
    \qedhere
\end{example}

The \emph{constituency parsing problem} states:
\begin{description}
    \item[Given:]
        a final state normalized CTA $\cA = (Q, \Sigma, \delta, q_f)$
        and
        $u \in (\Sigma^{(0)})^*$
    \item[Compute:] $\{ \xi \in \Lind(\cA) \mid \yield(\xi) = (u) \}$.
\end{description}
We note that, since $\cA$ is final state normalized,
every $\xi \in \Lind(\cA)$ has $\yield(\xi) \in \Sigma^*$.
Hence we did not allow string tuples consisting of more than one component
in the specification of the constituency parsing problem.

\section{Weighted RTG-based language models and the M-monoid parsing problem}
\label{sec:wrtglm}

The M-monoid parsing problem~\cite{moevog2019,moevog2021}
builds on RTG-based language models
which are inspired by
the initial algebra approach~\cite{Goguen1977}.

An \emph{RTG-based language model} (RTG-LM) is a tuple $(\cG,(\lalg{L},\phi))$ where,
for some $S$-signature $\Gamma$,
\begin{itemize}
    \item $(\lalg{L},\phi)$ is a $\Gamma$-algebra (\emph{language algebra}),
        we call the elements of~$\lalg{L}$ \emph{syntactic objects}, and
    \item $\cG=(N,\Lambda,A_0,R)$ is an $S$-sorted RTG with $\Lambda \subseteq \Gamma$.
\end{itemize}
The \emph{language generated by $(\cG, (\lalg{L},\phi))$} is the set
\[
    \lhom(\Lang(\cG)) = \{ \lhom(t) \mid t \in \Lang(\cG) \} \subseteq \lalg{L},
\]
i.e., the set of all syntactic objects obtained by evaluating trees of $\Lang(\cG)$ in the language algebra $\lalg{L}$.  
We note that $\lhom(\Lang(\cG)) \subseteq \lalg{L}^{\sort(A_0)}$,
i.e., each syntactic object in the language generated by $(\cG, (\lalg{L},\phi))$
has the sort of $A_0$.

A \emph{weighted RTG-based language model} (wRTG-LM) is a tuple
\[
    \big(\ (\cG,(\lalg{L},\phi)), \ \ (\walg{K},\oplus,\BZero,\Omega,\psi,\infsumop), \ \ \wt \ \big)
\]
where
\begin{itemize}
    \item $(\cG, (\lalg{L},\phi))$ is an RTG-LM,
    \item $(\walg{K},\oplus,\BZero,\Omega,\psi,\infsum)$ is a complete M-monoid (\emph{weight algebra}),
        and
    \item $\wt$ maps each rule of~$\cG$ with rank~$k$ to a $k$-ary operation in~$\Omega$.
        In the obvious way,
        we lift $\wt$ to the mapping $\wt': \T[R] \to \T[\Omega]$ and
        let $\wt$ also denote $\wt'$.
\end{itemize}

\noindent The \emph{M-monoid parsing problem} states:
\begin{description}
    \item[Given:] a wRTG-LM
        $((\cG,(\lalg{L},\phi)), (\walg{K},\oplus,\BZero,\Omega,\psi,\allowbreak\infsum), \allowbreak \wt)$
        with $G = (N,\Lambda,A_0,R)$
        and $a \in \lalg{L}$
    \item[Compute:] the value $\Mparse(a) \in \walg K$ where
        \[
            \Mparse(a) = \hspace{-1em}\infsum_{\substack{d \in \T[R]:\\\lhom(\projG(d))=a,\lhs(d(\varepsilon)) = A_0}}\hspace{-1em} \whom(\wt(d)).
        \]
\end{description}

\noindent
The computation of $\Mparse(a)$ employs the homomorphisms of both algebras.
Each AST of $\cG$ is mapped to an element of $\lalg{L}$
via the homomorphisms $\projG$ and $\lhom$
and it is mapped to an element of $\walg{K}$
via the homomorphisms $\wt$ and $\whom$.
Given a syntactic object $a$, the M-monoid parsing problem states to first compute
a collection of ASTs%
\footnote{Due to ambiguity, an AST may occur
several times in the computation of $\Mparse(a)$ \cite{moevog2021}.}
via the inverse of the homomorphisms $\projG$ and $\lhom$.
These ASTs are filtered for those where the left-hand side of the rule at the root
is the initial nonterminal.
Then, values in $\walg{K}$ are computed from the remaining ASTs
via the homomorphisms $\wt$ and $\whom$.
Finally, these values are accumulated to a single value using $\infsum$.

\section{Constituency parsing as an M-monoid parsing problem}
\label{sec:main}

In this section, we give the formal details of the definition of
the constituent tree algebra, the constituent tree yield algebra,
and the wRTG-LM we construct for a given CTA to model its constituency parsing problem.
Moreover, we sketch the proof of the statement that
the corresponding M-monoid parsing problem
is equal to that constituency parsing problem.
We start by defining partitioned constituent trees
which are inspired by the hybrid trees of Nederhof and Vogler~(2014)~\cite{nedvog14}
(also cf.~\cite{gebnedvog17,kuhnie2008}).

Let $\Sigma$ be a ranked alphabet.
A \emph{partitioned constituent tree (over $\Sigma$)} is
a tuple $\xi = (t, {<}, \allowbreak (U_1, \ldots, U_n))$ where
$t \in \T$,
$<$ is a strict total order on $\leaves(t)$,
$n \in \Nat_+$,
and
$(U_1, \ldots, U_n)$ is a partitioning of $\leaves(\xi)$ such that,
for every $i \in [n-1]$, $w_1 \in U_i$, and $w_2 \in U_{i+1}$, we have that $w_1 < w_2$.
Intuitively, this condition on the partitioning enforces consistency with $<$,
i.e., positions further left in $(U_1, \dots, U_n)$ are smaller.
We say that $\xi$ \emph{has $n$ segments}.
The set of all partitioned constituent trees over $\Sigma$ is denoted by $\pCT$.

Compared to the constituent trees of~\cite{dremoevog22},
partitioned constituent trees  abstract from particular indices.
Thus, each partitioned constituent tree represents infinitely many constituent trees.
To formalize this, we define the mapping $\rep\colon \CT \to \pCT$ as follows.
Let $\xi \in \CT$.
If $\xi$ is of the form $\ind{a}{n}$, we let
$\rep(\ind{a}{n}) = (a, \emptyset, (\{\varepsilon\}))$.
Otherwise, $\xi$ is
of the form $a(\xi_1, \ldots, \xi_k)$
and
we let $\rep(\xi) = (\projS(\xi), <, (U_1, \ldots, U_n))$ where,
for every $w_1, w_2 \in \leaves(\xi)$, we let $w_1 < w_2$ if and only if
$\projN(\xi(w_1)) < \projN(\xi(w_2))$ and $(U_1, \ldots, U_n)$ is the unique partitioning
of $\leaves(\xi)$ such that, for each $m \in [n]$, the set
$\{\projN(\xi(w)) \mid w \in U_m\}$ is an interval and, for each $m \in [n-1]$,
$\max_{w \in U_m} \projN(\xi(w)) + 1 < \min_{w \in U_{m+1}} \projN(\xi(w))$.
Intuitively, $<$ orders the leaves of $\xi$ by their indices and
$(U_1, \dots, U_n)$ groups the leaves such that, for each subset of the partitioning,
the indices of the leaves in that subset form an interval
and this interval is as large as possible.

We remark that our partitioned constituent trees
differ from the hybrid trees by~\cite{nedvog14} in three regards.
\begin{enumerate*}
    \item In the first component,
        we only allow a tree $\xi$ rather than a sequence of trees.
    \item The total order $<$ is defined on the set of leaves of
        $\xi$ rather than
        the set of all positions of $\xi$
        whose labels are from a particular subset $\Gamma$ of $\Sigma$.
        We note that~\cite{gebnedvog17} defined constituent trees%
        \footnote{They refer to constituent trees as phrase structure trees.}
        as a special case of hybrid trees where $\Gamma$ makes up the leaf labels,
        hence that difference is only syntactical
        (also, this was already indicated by~\cite{nedvog14}).
    \item Their hybrid trees did not feature a partitioning,
        so phrases with gaps cannot be modeled.
\end{enumerate*}
Compared to the segmented totally ordered terms (tots) of~\cite{kuhnie2008},
the total order of our partitioned constituent trees only regards the leaves
rather than the entire set of positions.

Intuitively, the linear phrase represented by a partitioned constituent tree
$(t, {<}, (U_1, \ldots, U_n))$ can be obtained
analogously to the yield of constituent trees in $\CT$;
we merely order the symbols at the leaves according to $<$ rather than by their index
and we place commas according to $(U_1, \ldots, U_n)$
rather than gaps in the indices.
We formalize this by defining the mapping
$\pyield\colon \pCT \to \Tup(\Sigma^*)$ as follows.
Let $\xi = (t, {<}, (U_1, \ldots, U_n))$ be in $\pCT$.
Then
\[
    \pyield(\xi) = (\pyaux(U_1), \ldots, \pyaux(U_n))
\]
where the auxiliary mapping $\pyaux$ is inductively defined by
$\pyaux(\emptyset) = \varepsilon$
and, for nonempty $U \subseteq \leaves(t)$,
$\pyaux(U) = t(\min\nolimits_< U) \cdot \pyaux(U \setminus \{ \min\nolimits_< U \})$.

It is easy prove that, for every $\xi \in \CT$, we have
\begin{equation}
    \label{eq-rep-preserves-yield}
    \yield(\xi) = \pyield(\rep(\xi)),
\end{equation}
i.e., intuitively, the mapping $\rep$ preserves $\yield$.

\subsection{The constituent tree algebra and the constituent tree yield algebra}

Prior to the definition of the algebras
we give the formal definition of their signature $\Gamma$.
The intuition behind our choice of sorts is the observation that
the elements of both algebras, partitioned constituent trees and string tuples,
have a certain \enquote{arity}:
each partitioned constituent tree has $n$ segments, i.e., groups of leaves,
and each string tuple consists of $n$ strings
where, in both cases, $n \in \Nat_+$.

We define the $((\Nat_+)^* \times \Nat_+)$-sorted set
$\Gamma = \Gamma^{(\varepsilon,1)} \cup \bigcup_{n,k,\ell_1,\ldots,\ell_k \in \Nat_+} \Gamma^{(\ell_1\cdots\ell_k,n)}$ where
\begin{itemize}
    \item $\Gamma^{(\varepsilon, 1)} = \Sigma^{(0)}$
        and
    \item for every $n, k, \ell_1, \ldots, \ell_k \in \Nat_+$, we let
        \[
            \Gamma^{(\ell_1 \cdots \ell_k, n)} = \{ (a, e) \mid
                a \in \Sigma^{(k)}, e \in \W[n][(\ell_1, \ldots, \ell_k)] \}.
        \]
\end{itemize}

Now we can approach the definition of the constituent tree algebra
as a $\Gamma$-algebra whose carrier set is $\pCT$.
For this, we consider $\pCT$ as an $\Nat_+$-sorted set
by letting, for every $n \in \Nat_+$,
\[
    (\pCT)^{(n)} = \{ \xi \in \pCT \mid \text{$\xi$ has $n$ segments} \}.
\]
The \emph{constituent tree algebra} is the $\Nat_+$-sorted $\Gamma$-algebra
$\CTAlg = (\pCT, \CTConst)$ where
\begin{itemize}
    \item for each $a \in \Sigma^{(0)}$, we let
        $\CTConst{a} = (a, \emptyset, (\{\varepsilon\}))$ and
    \item for every
        $(a,e) \in \Gamma^{(\ell_1 \cdots \ell_k, n)}$
        and 
        $\xi_1 \in (\pCT)^{(\ell_1)}, \allowbreak \ldots, \allowbreak \xi_k \in (\pCT)^{(\ell_k)}$
        with $\xi_i = (t_i, {<_i}, (U_1^{(i)}, \dots, U_i^{(\ell_i)}))$
        (for $i \in [k]$),
        we let
        \[
            \CTConst{a, e}(\xi_1, \ldots, \xi_k)
            = (t, <, (U_1, \ldots, U_n))
        \]
        where $t = a(t_1, \ldots, t_k)$ and, for each $m \in [n]$, we let
        $U_m$ be the union of all sets $\{i\} \cdot U_i^{(j)}$
        such that $x_i^j$ occurs in the $m$-th component of $e$.
        Thus, clearly, $(U_1, \dots, U_n)$ is a partitioning of $\leaves(t)$.
        Hence, for each $w \in \leaves(t)$, there exist exactly one
        $i \in [k]$ and $j \in [\ell_i]$ such that $w \in \{i\} \cdot U_i^{(j)}$;
        we let $\var(w)$ denote $x_i^j$.
        For the definition of $<$,
        let $w_1, w_2 \in \leaves(t)$.
        If $\var(w_1) \not= \var(w_2)$, then we let $w_1 < w_2$ if and only if
        $\var(w_1)$ occurs left of $\var(w_2)$ in $e$.
        Otherwise, we let $i \in [k]$ and $j \in [\ell_i]$ such that
        $\var(w_1) = x_i^j$.
        Then $w_1 < w_2$ if and only if $w_1' <_i w_2'$
        where $w_1', w_2' \in \pos(t_i)$ such that
        $w_1 = i \cdot w_1'$ and $w_2 = i \cdot w_2'$.
\end{itemize}
We let $\ctahom$ denote the unique $\Gamma$-homomorphism from $\T[\Gamma]$ to $\pCT$.

We note that the definition of $\CTAlg$
is semantically close to the algebra of segmented tots by~\cite{kuhnie2008},
but the operations of $\CTAlg$ are defined using word tuples
and the non-nullary symbols of $\Gamma$ do not add tree positions
to the total order or the partitioning since, in our case,
these components only refer to the leaves.
Moreover, one cannot define a $\Gamma$-algebra similar to $\CTAlg$ but with $\CT$
as its carrier set.
For this, one would need to fix a mapping $\sort: \CT \to \Nat_+$.
An appropriate choice could be assigning to each $\xi \in \CT$
the smallest number $n$ such that the indices of $\xi$ form $n$ intervals.
For instance, let $\xi_1 = \ind{a}{2}$ and $\xi_2 = b(\ind{a}{1}, \ind{a}{4})$
be constituent trees over $\Sigma$.
Then we have $\sort(\xi_1) = 1$ and $\sort(\xi_2) = 2$.
In essence, this mimics the sort mapping of $\pCT$ but considers intervals of indices
rather than the partitioning of the set of leaves.
However, this approach bears the following problem.
Let $c \in \Sigma^{(2)}$ and $e = (x_2^1 x_1^1, x_2^2)$.
We compute $\CTConst(c, e)(\xi_1, \xi_2) = a(\xi_1, \xi_2)$ and have
$\sort(a(\xi_1, \xi_2)) = 2$.
On the other hand, if we also consider $\xi_3 = b(\ind{a}{1}, \ind{a}{3})$,
then
$\CTConst(c,e)(\xi_1,\xi_3)$
has sort $1$ which contradicts the
sort of $(c, e)$.
Moreover, this sort mapping falls short of inhibiting that constituent trees
with overlapping indices are passed as arguments to $\CTConst(e)$.
The rich field of many-sorted algebra surely provides means to
remedy these problems by choosing a more complex signature rather than $\Gamma$.
However, we believe that circumventing these problems by dealing with $\pCT$
is a cleaner solution.

We define the \emph{constituent tree yield algebra} to be
the $\Nat_+$-sorted $\Gamma$-algebra $(\Tup(\Sigma^*), \CTString)$ where
\begin{itemize}
    \item for each $n \in \Nat_+$, we let $\sort(\Tup[n](\Sigma^*)) = n$,
    \item for each $a \in \Sigma^{(0)}$, we let $\CTString{a} = (a)$, and
    \item for each $(a,e) \in \Gamma^{(\ell_1 \cdots \ell_k, n)}$,
        we let $\CTString{a,e} = \sem(e)$.
\end{itemize}
Let $\lcfrshom$ denote the unique homomorphism from $\T[\Gamma]$ to $\Tup(\Sigma^*)$.
We can also show that the mapping $\pyield$ is
a $\Gamma$-homomorphism from $\pCT$ to $\Tup(\Sigma^*)$.
Thus, by the laws of universal algebra (cf., e.g., \cite{wec92}),
we obtain that, for every $t \in \T[\Gamma]$,
\begin{equation}\label{eq-ctalg-yield-homomorphism}
    \pyield(\ctahom(t)) = \lcfrshom(t).
\end{equation}

\subsection{The wRTG-LM for constituency parsing}

Here we show, given a CTA $\cA$ and a string $u$,
how to construct a wRTG-LM such that the corresponding M-monoid parsing problem
is equivalent to the constituency parsing problem for $\cA$ and $u$.
We start with the RTG and afterwards add the algebras from the previous subsection.

Let $\cA = (Q, \Sigma, \delta, q_f)$ be a CTA.
We define the \emph{$\cA$-RTG} to be the RTG $\cG = (Q, \Lambda, R, q_f)$ where
\begin{enumerate}
    \item $\Lambda = \Lambda^{(\varepsilon,1)} \cup \bigcup_{k\in\Nat_+,n,\ell_1,\ldots,\ell_k \in \rk(Q)} \Lambda^{(\ell_1\cdots\ell_k,n)}$ where
        we let
        $\Lambda^{(\varepsilon,1)} = \Sigma^{(0)}$ and,
        for each $k \in \Nat_+$ and every $n, \ell_1, \dots, \ell_k \in \rk(Q)$,
        we let
        $\Lambda^{(\ell_1\cdots\ell_k,n)} = \Sigma^{(k)} \times \W[n][(\ell_1,\dots,\ell_k)]$,%
        \footnote{Thus $\Lambda$ is a finite subset of $\Gamma$.}
    \item for every $a \in \Sigma^{(0)}$ and $q \in Q$ it holds that
        $(\varepsilon, a, q) \in \delta$ if and only if
        $(q \to (a)) \in R$, and
    \item for every $k \in \Nat_+$, $a \in \Sigma^{(k)}$, $e \in \WE$, and
        $q_1, \ldots, q_k, q \in Q$ it holds that $(q_1 \ldots q_k, a, e, q) \in \delta$
        if and only if $(q \to (a, e)(q_1, \ldots, q_k)) \in R$.
\end{enumerate}

\begin{example}\label{ex:wrtglm}
    Recall the CTA $\cA = (Q, \Sigma, \delta, q_f)$ from Example~\ref{ex:cta}.
    The $\cA$-RTG is $\cG = (Q, \Lambda, R, q_f)$
    where $\Lambda^{(\varepsilon,1)} = \{a,b,c\}$,
    for every $n,\ell_1,\ell_2,\ell_3 \in [3]$,
    $\Lambda^{(\ell_1\ell_2,n)} = \{e\} \times \W[n][(\ell_1,\ell_2)]$ and
    $\Lambda^{(\ell_1\ell_2\ell_3,n)} = \{d\} \times \W[n][(\ell_1,\ell_2,\ell_3)]$;
    and $R$ consists of the following rules:
    \[
        \begin{gathered}
            q_f \to (d, (x_1^1 x_2^1 x_1^2 x_2^2 x_3^1 x_2^3))(q_l, q, q_c) \qquad
            q_f \to (d, (x_1^1 x_2^1 x_3^1 x_2^2 x_3^2 x_2^3))(q_a, q, q_r) \\
            q \to (d, (x_1^1 x_2^1, x_1^2 x_2^2, x_3^1 x_2^3))(q_l, q, q_c) \qquad
            q \to (d, (x_1^1 x_2^1, x_3^1 x_2^2, x_3^2 x_2^3))(q_a, q, q_r) \\
            q \to (d, (x_1^1, x_2^1, x_3^1))(q_a, q_b, q_c) \qquad
            q_l \to (e, (x_1^1, x_2^1))(q_a, q_b) \qquad
            q_r \to (e, (x_1^1, x_2^1))(q_b, q_c) \\
            q_a \to (a) \qquad
            q_b \to (b) \qquad
            q_c \to (c).
        \end{gathered}
    \]
    The bottom right of Figure~\ref{fig:main} shows an AST of $\cG$.
\end{example}

As the language algebra of our wRTG-LM
we use the constituent tree yield algebra $(\Tup(\Sigma^*), \CTString)$.
For the weight algebra we point out that each of its operations
computes a single partitioned constituent tree.
However, our goal as determined by the constituency parsing problem
is to compute a \emph{set} of constituent trees.
Hence, we lift the constituent tree algebra to sets.
Formally, we define the M-monoid
\def\natcap{\sortedcup[\scalebox{0.9}{$\Nat$}]}%
\[
    \CTparse = (\bigcup_{n \in \Nat_+} \PowerSet{\pCT^{(n)}} \cup \{\bot\}, \natcap, \emptyset, \Gamma, \CTConstp, \infsumop[\natcap])
\]
where $\natcap$ and $\CTConstp$ are defined like
their counterparts in Example~\ref{ex:mmonoid}.
In the following, we will write $\CTConst$ rather than $\CTConstp$
and we let $\ctahom$ denote also the unique $\Gamma$-homomorphism from $\T[\Gamma]$
to this algebra.

Combining these components,
we define the \emph{$\cA$-wRTG-LM} to be the wRTG-LM
\[
    \bar{G} = ((\cG, (\Tup(\Sigma^*), \CTString)), \CTparse, \wt)
\]
where $\cG$ is the $\cA$-RTG
and, for every $r = (A \to \gamma(A_1, \ldots, A_k))$ in $R$, we let $\wt(r) = \gamma$.

\subsection{Constituency parsing is an instance of the M-monoid parsing problem}

Let $\cA = (Q, \Sigma, \delta, q_f)$ be a CTA and
let $\bar{G} = ((\cG, (\Tup(\Sigma^*), \CTString)), \CTparse, \wt)$
with $\cG = (Q, \Lambda, R, q_f)$
be the $\cA$-wRTG-LM.
The M-monoid parsing problem for $\bar{G}$ is, given some $(u) \in \Tup(\Sigma^*)$,
to compute
\[
    \Mparse[(\cG,\CT)](u) = \bigcup_{\substack{d \in \T[R]:\\
    \lcfrshom(\projG(d)) = (u), \, \lhs(d(\varepsilon)) = q_f}}
    \ctahom(\wt(d)).
\]
For a given phrase $u$, this instance of the M-monoid parsing problem enumerates
the set of all ASTs of $\cG$ that have the initial nonterminal at the root
and evaluate to $u$ in the constituent tree yield algebra.
Each of these ASTs is evaluated in the constituent tree algebra.

In order to show that
this M-monoid parsing problem is equal to
the constituency parsing problem for $\cA$ (and $u$),
we seek a bijection $\bij$ between the set $\LR$
and the set of abstract syntax trees of $\cG$.
However, similar to~\cite{dremoevog22}, we only find such a bijection
if we consider certain elements of $\LR$ equivalent.

Formally, we define the equivalence relation $\eqrel$ as follows.
For every $(\xi_1, \rho_1), (\xi_2, \rho_2) \in \LR$, we let
$(\xi_1, \rho_1) \eqrel (\xi_2, \rho_2)$ if and only if
$\projS(\xi_1) = \projS(\xi_2)$ and $\rho_1 = \rho_2$.
Clearly, $\eqrel$ is indeed an equivalence relation.
Let $\quot{\LR}$ denote the quotient set of $\LR$ by $\eqrel$.
For each $(\xi, \rho) \in \LR$, we let $\eqcl{\xi,\rho}$ denote
the equivalence class $(\xi,\rho)$ belongs to.
An example for $\eqrel$ is given in the top of Figure~\ref{fig:main}.

We define the mapping $\bij\colon \quot{\LR} \to \T[R]$ inductively as follows.
Let $(\xi, \rho) \in \LR$.
If $(\xi, \rho)$ is of the form $(\ind{a}{n}, q)$, we let
$\bij(\eqcl{\ind{a}{n}, q}) = q \to (a)$.
Otherwise,
$\xi$ is of the form $a(\xi_1, \ldots, \xi_k)$ and
$\rho$ is of the form $(q, e)(\rho_1, \ldots, \rho_k)$,
then we let
\[
    \bij(\eqcl{\xi, \rho}) = q \! \to \!(a, e)(\bij(\eqcl{\xi_1, \rho_1}), \ldots, \bij(\eqcl{\xi_k, \rho_k})).
\]
We illustrate $\bij$ for the CTA $\cA$ from Example~\ref{ex:cta}
and the $\cA$-RTG $\cG$ from Example~\ref{ex:wrtglm} in
Figure~\ref{fig:main}.

Using a method similar to~\cite{dremoevog22} we can show that
$\bij$ is indeed a bijection.
Moreover, we can prove the following auxiliary statement.
Let $(\xi, \rho) \in \LR$.
If $\ctahom(\projG(\bij(\eqcl{\xi, \rho}))) = (t_1, <_1, (U_1^{(1)}, \ldots, U_1^{(\ell_1)}))$ and
$\rep(\xi) = (t_2, {<_2}, (U_2^{(1)}, \ldots, U_2^{(\ell_2)}))$,
then
\begin{equation}\label{eq-bij-equal-rep}
    t_1 = t_2 \quad \text{and} \quad {<_1} = {<_2}.
\end{equation}
Intuitively, $\ctahom(\projG(\bij(\eqcl{\xi, \rho})))$ and $\rep(\xi)$ may only differ
in the partitioning.
For instance,
consider the constituent tree $\xi$ and the run $\rho$ in Figure~\ref{fig:main}
where we even have $\ctahom(\projG(\bij(\eqcl{\xi, \rho}))) = \rep(\xi)$.

We note that since $\cA$ is final state normalized,
$\bij$ implies that each AST $d$ of $\cG$ with $\lhs(d(\varepsilon)) = q_f$
has $\lcfrshom(\projG(d)) \in \Sigma^*$.
Thus, $\Mparse[(\cG,\CT)](u)$ is only non-empty if $u$ is a string.
This resembles the fact that the constituency parsing problem is only defined for strings.
We will assume that $u \in \Sigma^*$ in the following.

After these preparations, we can show that the M-monoid parsing problem
for $\bar{G}$ and $u$
relates to the constituency parsing problem for $\cA$ and $u$
in the following way:
\begingroup
\allowdisplaybreaks
\begin{align*}
    \Mparse[(\cG,\CT)](u)
    &= \{ \ctahom(\wt(d)) \mid d \in \T[R], \lcfrshom(\projG(d)) = u, \lhs(d(\varepsilon)) = q_f \} \\
    &\stackrel{\eqref{eq-ctalg-yield-homomorphism}}=
    \{ \ctahom(\wt(d)) \mid d \in \T[R], \pyield(\ctahom(\projG(d))) = u, \lhs(d(\varepsilon)) = q_f \} \\
    &\stackrel{\text{bij.}}= \{ \ctahom(\wt(\bij(\eqcl{\xi, \rho}))) \mid
        \begin{aligned}[t]
            &(\xi, \rho) \in \LR, \pyield(\ctahom(\projG(\bij(\eqcl{\xi, \rho})))) = u, \\
            &\lhs(\bij(\eqcl{\xi, \rho})(\varepsilon)) = q_f \}
        \end{aligned}\\
    &\stackrel{\star_1}= \{ \ctahom(\wt(\bij(\eqcl{\xi, \rho}))) \mid
        \begin{aligned}[t]
            &(\xi, \rho) \in \LR, \pyield(\ctahom(\projG(\bij(\eqcl{\xi, \rho})))) = u, \\
            &\text{$q_f$ is the state at $\rho(\varepsilon)$} \}
        \end{aligned}\\
    &\stackrel{\star_2}=
    \{ \rep(\xi) \mid (\xi, \rho) \in \LR, \pyield(\rep(\xi)) = u,
            \text{$q_f$ is the state at $\rho(\varepsilon)$} \}
        \\
    &\stackrel{\text{\eqref{eq-rep-preserves-yield}}}=
    \{ \rep(\xi) \mid (\xi, \rho) \in \LR, \yield(\xi) = u,
        \text{$q_f$ is the state at $\rho(\varepsilon)$} \} \\
    &= \{ \rep(\xi) \mid \xi \in \Lind(\cA), \yield(\xi) = u \}
\end{align*}
\endgroup
where $\star_1$ holds by definition of $\bij$ and
$\star_2$ follows from~\eqref{eq-bij-equal-rep} (using $\wt = \projG$) and
both $\rep(\xi)$ and $\ctahom(\wt(\bij(\eqcl{\xi,\rho})))$ being in $\pCT^{(1)}$
(which is a consequence of $\cA$ being final state normalized).
%
We illustrate this equality by showing how
the mapping $\rep$ commutes with $\ctahom \circ \projG \circ \bij$
in Figure~\ref{fig:main}.

We note that $\Mparse[(\cG,\CT)](u)$ is a subset of $\pCT$
(i.e., constituent trees without particular indices)
whereas the constituency parsing problem computes a subset of $\CT$.
To bridge this gap, we note that 
the set $T = \{ \xi \in \CT \mid \rep(\xi) \in \Mparse[(\cG,\CT)](u) \}$
can be easily constructed.
We sketch this construction by letting
$\xi = (t, {<}, (U_1, \ldots, U_n)) \in \Mparse[(\cG,\CT)](u)$.
Since $\cA$ is final state normalized, we have $n = 1$.
Now we let $m \in \Nat_+$ and fix an interval $[m, m + |U_1|]$,
then we obtain $\xi' \in \CT$
from $t$ by adding the indices $m, m+1, \ldots, m + |U_1|$ to the symbols at
the leaves of $t$ in the order determined by $<$.
Clearly, $\rep(\xi') = \xi$.
By letting $m$ range over $\Nat_+$ we obtain
the set $\{ \xi' \in \CT \mid \rep(\xi') = \xi \}$.
Clearly, for every $\xi \in T$ we have $\yield(\xi) = u$ and $\xi \in \Lind(\cA)$.
Thus $T$ is the solution of the constituency parsing problem of $\cA$ and $u$.

%

\section{Applicability of the M-monoid parsing algorithm}
\label{sec:algorithm}

The M-monoid parsing algorithm~\cite{moevog2021} is a two-phase pipeline
which is applicable to a large class of M-monoid parsing problems,
where applicability means that the algorithm is terminating and correct.
Due to space restrictions, we cannot repeat the algorithm here and
only investigate its applicability to our scenario.
For this, we let $\WRTGLM(\CTA)$ be the set of all $\cA$-wRTG-LMs
for each final state normalized CTA $\cA$.
We let $\bar{G} \in \WRTGLM(\CTA)$ and $u \in \Sigma^*$.

The first phase of the M-monoid parsing algorithm applies a weighted deduction system
to $\bar{G}$ and $u$, thus obtaining a new wRTG-LM $\bar{G'}$.
Mörbitz and Vogler~(2021)~\cite{moevog2021} provide the canonical weighted deduction system which is applicable
in all situations where the language algebra of the input wRTG-LM is
finitely decomposable.
Since this is clearly the case for $(\Tup(\Sigma^*), \CTString)$,
we obtain that the first phase of the M-monoid parsing algorithm
is applicable to every $\bar{G} \in \WRTGLM(\CTA)$ and $u \in \Sigma^*$.

The second phase, called value computation algorithm,
uses $\bar{G'}$ to compute an element in the weight algebra.
There are two independent sufficient conditions for this value
to be equal to $\Mparse[(\cG,\CT)](u)$.
The first condition requires $\bar{G}$ to fulfil a property called closed.
Without giving details on this property, we state that not every
wRTG-LM in $\WRTGLM(\CTA)$ is closed.%
\footnote{The interested reader may see that for themselves in a way
similar to~\cite[Appendix A.7]{moevog2021}.}
The second condition requires $\bar{G}$ to fulfil a property called nonlooping
and the weight algebra to be distributive and d-complete.
Now distributivity of $\CTparse$ is easy to see and d-completeness of $\CTparse$
follows from the fact that its additive monoid is completely idempotent.
In essence, $\bar{G}$ is nonlooping if for each AST $d$ of its RTG the following holds:
if there is a proper subtree $d|_w$ of $d$
which evaluates to the same syntactic object as $d$ in the language algebra,
then $d(w)$ must have a different label than $d(\varepsilon)$.
As our language algebra is $(\Tup(\Sigma^*), \CTString)$,
this property can only be violated if each node in $d$ from $\varepsilon$ to $w$
is monadic.
Then, by pumping the monadic chain from $\varepsilon$ to $w$,
we can construct infinitely many ASTs with the same yield,
each of which is evaluated to a different constituent tree in the weight algebra.
However, a terminating algorithm cannot compute an infinite set of constituent trees.
By the construction of $\bar{G}$, we find that this situation is only possible
if the CTA $\cA$ contains transitions of the form $(q_1, a_1, e_1, q_2)$,
$(q_2, a_2, e_2, q_3)$, \dots, $(q_n, a_n, e_n, q_1)$.
Thus, if $\cA$ is free of such monadic cycles,
$\bar{G}$ is nonlooping and
the M-monoid parsing algorithm is correct for $\bar{G}$ and~$u$.

\section{Future work}

Dependency is another important syntactical analysis in NLP.
Dependency trees are also introduced by~\cite{dremoevog22}
where dependency tree automata are mentioned as another possible special case of HTA,
mirroring CTA.
We believe that the corresponding dependency parsing problem
can be shown to be an instance of M-monoid parsing
in a way very similar to the present paper.

The constituency parsing problem considered here states to compute the set of
all suitable constituent trees.
However, parsing problems often occur in weighted settings
where the weights are, e.g., probabilities, and compute only the best analysis.
A constituency parsing problem with such additional weights
also falls in the scope of the M-monoid parsing problem.
Moreover, the underlying CTA could even have transitions that allow monadic cycles
as long as they lead to a decrease in weight.

\bibliographystyle{eptcsini}
\bibliography{lit.bib}

\end{document}

%% file: math-notation.tex
\mathchardef\mhyphen="2D 

\newcommand*\alg[1]{\mathcal{#1}}
\newcommand*\lalg[1]{\alg{#1}}
\newcommand*\walg[1]{\mathbb{#1}}
\newcommand*\welem[1]{\mathbb{#1}}

\newcommand*\Nat{\mathbb N}

\newcommand*\ints{\mathbb I}
\NewDocumentCommand\Vars{o}{\mathbb X\IfNoValueTF{#1}{}{_{#1}}}
\newcommand*{\PowerSet}[1]{\mathcal{P}(#1)}
\newcommand*\T[1][\Sigma]{\mathrm{T}_{#1}}

\newcommand*\pCT[1][\Sigma]{\mathrm{pC}_{#1}}

\newcommand*\CT[1][\Sigma]{\mathrm{C}_{#1}}
\newcommand*\BZero{\mathbb{0}}

\newcommand*\CTAlg{\alg{CT}}
\NewDocumentCommand\CTConst{g}{\theta_\Sigma\IfNoValueTF{#1}{}{(#1)}}
\NewDocumentCommand\CTConstp{g}{\theta_\Sigma'\IfNoValueTF{#1}{}{(#1)}}
\NewDocumentCommand\CTString{g}{\theta_{\mathrm{Y}}\IfNoValueTF{#1}{}{(#1)}}
\newcommand*\Var[1][\kappa]{\mathbb{X}_{#1}}
\newcommand*\ICAll{\mathrm{UIC}}
\newcommand*\ICComponent{\mathrm{ic}}
\newcommand*\WE{\mathbb W}
\NewDocumentCommand\W{O{n}O{\kappa}}{\WE^{#1}_{#2}}

\newcommand*\runs[2][]{\mathrm{R}_{#2}^{#1}}

\newcommand*\indset[1][\cA]{\ensuremath{\Uptheta}_{#1}}
\newcommand*\indrep[1][\cA]{\ensuremath{\Uptheta}_{#1}'}
\newcommand*\LR[1][\cA]{\mathrm{CR}_{#1}}
\newcommand*\RT[1][\cG]{\mathrm{RT}_{#1}}
\newcommand*\CTparse{\mathbb C}

\newcommand*\cA{\mathcal A}
\newcommand*\cG{\mathcal G}
\newcommand*\cW{\mathcal W}

\newcommand*\ind[2]{#1\langle #2\rangle}

\NewDocumentCommand\NewMapping{s o m}%
 {\expandafter\NewDocumentCommand\csname#3\endcsname{d()}%
 {\IfBooleanTF{#1}%
 {(\IfNoValueTF{##1}{\cdot}{##1})_{#2}}%
 {\IfNoValueTF{#2}{\mathrm{#3}}{#2}\IfNoValueTF{##1}{}{(##1)}}}}%
\NewMapping{fo}
\NewMapping{rep}
\NewMapping[\psi]{bij}
\NewMapping[\#_{\mathrm{lv}}]{lv}
\NewMapping[\varphi]{ass}
\NewMapping{var}
\NewMapping{yield}
\NewMapping[\mathrm{p\mhyphen{}yield}]{pyield}
\NewMapping[\mathrm{yield}']{yieldP}
\NewMapping[\mathrm{I\mhyphen{}yield}]{Iyield}
\NewMapping[\mathrm{split}]{Isplit}
\NewMapping[\mathrm{cc}]{concat}
\NewMapping[\mathrm{I\mhyphen{}sort}]{Isort}
\NewMapping{sort}
\NewMapping[\mathrm{L}]{Lang}
\NewMapping[\mathrm{L}_{\mathrm{ind}}]{Lind}
\NewMapping{wt}
\NewMapping{lhs}
\NewMapping{cwds}
\NewMapping{leaves}
\NewMapping{pos}
\NewMapping{AST}
\NewMapping{factors}
\NewMapping[\mathrm{id}]{idm}
\NewMapping{seq}
\NewMapping{mul}
\NewMapping{ic}
\NewDocumentCommand\iass{o O{\xi}O{\rho}}{\indrep(#2\IfNoValueF{#1}{|_{#1}},#3\IfNoValueF{#1}{|_{#1}})}
\NewDocumentCommand\sem{d()}{\llbracket\IfNoValueTF{#1}{\cdot}{#1}\rrbracket}
\NewDocumentCommand\semP{d()}{\llbracket\IfNoValueTF{#1}{\cdot}{#1}\rrbracket'}
\NewDocumentCommand\NewMappingWithOpt{s o o m}%
 {\expandafter\NewDocumentCommand\csname#4\endcsname{o d()}%
 {\IfBooleanTF{#1}%
 {(\IfNoValueTF{##2}{\cdot}{##2})_{#2}}%
 {\IfNoValueTF{#2}{\mathrm{#4}}{#2}%
  \IfValueTF{##1}{_{##1}}{\IfValueT{#3}{_{#3}}}\IfNoValueTF{##2}{}{(##2)}}}}%
\NewMappingWithOpt{rk}
\NewMappingWithOpt[\mathrm{parse}][(\cG, \alg{L})]{Mparse}
\NewMappingWithOpt[f][t,{<}]{pyaux}
\NewMappingWithOpt{Tup}

\newcommand{\infsum}[1][\oplus]{\mathop{\vphantom{\sum}\mathchoice
  {\vcenter{\hbox{\Large $\sum^{#1}$}}}
  {\vcenter{\hbox{\normalsize $\sum^{#1}$}}}{\sum^{#1}}{\sum^{#1}}}\displaylimits}
\newcommand{\infsumop}[1][\oplus]{\mathop{\vphantom{\sum}\mathchoice
  {\vcenter{\hbox{\normalsize $\sum^{#1}$}}}
  {\vcenter{\hbox{\normalsize $\sum^{#1}$}}}{\sum^{#1}}{\sum^{#1}}}\displaylimits}
\newcommand*\sortedcup[1][S]{\stackinset{c}{}{c}{.5ex}{\scriptsize\ensuremath{#1}}{$\cup$}}

\NewDocumentCommand\proj{m d()}{(\IfNoValueTF{#2}{\cdot}{#2})_{#1}}
\NewDocumentCommand\NewProjection{o m}%
 {\expandafter\newcommand\csname#2\endcsname{\proj#1}}%
\NewProjection[\Sigma]{projS}
\NewProjection[\Nat]{projN}
\NewProjection[Q]{projQ}
\NewProjection[\WE]{projW}
\NewProjection[\ICComponent]{projIC}
\NewProjection[\Gamma]{projG}

\NewMapping*[\lalg{L}]{lhom}
\NewMapping*[\walg{K}]{whom}
\NewMapping*[\CTAlg]{ctahom}
\NewMapping*[\mathrm{Y}]{lcfrshom}

\newcommand*\ltfac{\mathrel{<_{\mathrm{factor}}}}
\newcommand*\eqrel{\sim}
\newcommand*\quot[1]{{#1}/_\eqrel}
\newcommand*\eqcl[1]{[{#1}]}

\newcommand*\CTA{\ensuremath{\mathrm{CTA}}}
\NewMapping[\cW]{WRTGLM}

[style=definition]
[sibling=theorem,style=definition]
[sibling=theorem,style=definition]
[sibling=theorem,style=definition]
\declaretheorem{example}[sibling=theorem,style=definition,qed={$\triangleleft$}]

%% file: fig1.tex
\begin{tikzpicture}[remember picture, scale=0.73, every node/.style={transform shape},
level distance=0.9cm,sibling distance=.8cm]

\begin{scope}[execute at end node={\vphantom{Vg}},
              level 1/.style={sibling distance=23.5mm},
              level 2/.style={sibling distance=14mm},
              edge from parent path={(\tikzparentnode.south) -- +(0,-4pt) -| (\tikzchildnode)}]
\node (n) {VP}
  child { node (n1) {V}
    child { node (n11) {hat} }
    child { node (n12) {gearbeitet} } }
  child { node (n2) {ADV}
    child { node (n21) {schnell} } };

\node[yshift=-1.1cm] (hat) at (n11) {hat};
\node[yshift=-1.1cm] (schnell) at (n12) {schnell};
\node[yshift=-1.1cm] (gearbeitet) at (n21) {gearbeitet};
\draw[densely dashed] (n11) -- (hat);
\draw[densely dashed] (n12) -- (gearbeitet);
\draw[densely dashed] (n21) -- (schnell);
\end{scope}

\begin{scope}[execute at end node={\vphantom{Vg}}, xshift=6.25cm,
              level 1/.style={sibling distance=31mm},
              level 2/.style={sibling distance=18mm},]
\node (x) {VP}
  child { node (x1) {V}
    child { node (x11) {hat$\langle 1\rangle$} }
    child { node (x12) {gearbeitet$\langle 3\rangle$} } }
  child { node (x2) {ADV}
    child { node (x21) {schnell$\langle 2\rangle$} } };
\node[above left=-0.3cm and 1cm of x] {\strut $\xi :$};

\draw[gray] ([yshift=2.5mm]x1.north) [rounded corners = 6mm] -- ([xshift=-5mm]x11.south west) [rounded corners = 9mm] -- ([xshift=7mm]x12.south east) [rounded corners = 3mm] -- cycle;

\node[yshift=-1.1cm, anchor=west] (yield) at (x11.west) {$\text{yield}(\xi)= ($hat schnell gearbeitet$)$};
\node[yshift=-.6cm,anchor=west,gray] at (yield.west) {$\text{yield}(\xi |_1)= ($hat, gearbeitet$)$};
\end{scope}

\begin{scope}[execute at end node={\vphantom{Vg$x_1^1$}}, xshift=15cm,
              level 1/.style={sibling distance=50mm},
              level 2/.style={sibling distance=30mm},]
\node (d) {VP $\to (\text{VP}, (x_1^1 \, x_2^1 \, x_1^2))(\text{V}, \text{ADV})$}
  child { node (d1) {V $\to (\text{V}, (x_1^1 , x_2^1))(\text{h}, \text{g})$}
    child { node (d11) {h $\to (\text{hat})$} }
    child { node (d12) {g $\to (\text{gearbeitet})$} } }
  child { node (d2) {ADV $\to (\text{ADV}, (x_1^1))(\text{s})$}
    child { node (d21) {s $\to (\text{schnell})$} } };    
\node[above left= -0.3cm and 0.9cm of d] {\strut $d:$};
\end{scope}

\draw[thick, rounded corners=4mm, ->] ([shift=(135:.7cm)]d.center) -- ([shift=(135:1.3cm)]d.center) --  node[pos=0.4, above] {$\ctahom \circ \projG$} ([shift=(45:1.3cm)]n.center)  -- ([shift=(45:.7cm)]n.center);
\draw[thick, rounded corners=6mm, ->] ([yshift=-3mm]d12.south) -- ([xshift=4.75cm]yield.east) -- node[pos=0.3, below] {$\lcfrshom \circ \wt$} ([xshift=3mm]yield.east);
\end{tikzpicture}

%% file: fig2-tighter.tex
\begin{tikzpicture}[remember picture, scale=0.75, node distance= -5mm and 0.5cm,
                    yscale=0.8, myyscale/.style={yscale=0.8},
                    every node/.style={transform shape,yscale=1/0.8},
                    execute at end node={\strut},
                    myfill/.style={fill=white,rounded corners=1mm,inner xsep=0.8mm,outer xsep=0.2mm, inner ysep=-0.1mm, outer ysep=2.1mm}] 

\begin{scope}[xshift=0.8mm, yshift=-3.5mm, mygray,
              inner ysep=0.4mm, outer ysep=0mm,
              level distance= 1.5cm,
              level 1/.style={sibling distance=29mm},
              level 2/.style={sibling distance=20mm},
              level 3/.style={sibling distance=11mm}]
\node (xi2root) {\phantom{$d$}}
  child[level distance= 1cm]{ node[inner ysep=-1mm, outer ysep=0mm,inner xsep=3mm] (xi2n1) {\vphantom{$e$}} 
    child[level distance= 0.9cm,sibling distance=11mm]{ node {} }
    child[level distance= 0.9cm,sibling distance=11mm]{ node {} } }
  child[xshift=0mm]{ node (xi2n2) {\phantom{$d$}} 
    child[xshift=6mm,level distance= 1.35cm]{ node {} }
    child{ node (xi2n21) {\phantom{$d$}} 
      child{ node {} }
      child{ node {} }
      child{ node {} }  }
    child[xshift=5mm,level distance= 0.9cm]{ node[inner ysep=-1mm, outer ysep=0mm,inner xsep=3mm] (xi2n22) {\vphantom{$e$}}
      child{ node {} }
      child{ node {} } } }
  child[level distance= 1cm]{ node {} };
\end{scope}

\begin{scope}[every node/.append style={myfill},
              level distance= 1.5cm,
              level 1/.style={sibling distance=29mm},
              level 2/.style={sibling distance=20mm},
              level 3/.style={sibling distance=11mm}]
\node (xiroot) {$d$}
  child[level distance= 1cm]{ node {$e$} 
    child[level distance= 0.9cm,sibling distance=11mm]{ node[inner ysep=-1mm] (xil1) {$\ind{a}{1}$} }
    child[level distance= 0.9cm,sibling distance=11mm]{ node[inner ysep=-1mm] (xil2) {$\ind{b}{4}$} } }
  child[xshift=0mm,myfill]{ node {$d$} 
    child[xshift=6mm,level distance= 1.35cm]{ node (xil3) {$\ind{a}{2}$} }
    child{ node {$d$} 
      child{ node (xil4) {$\ind{a}{3}$} }
      child{ node (xil5) {$\ind{b}{6}$} }
      child{ node (xil6) {$\ind{c}{9}$} }  }
    child[xshift=5mm,level distance= 0.9cm]{ node {$e$}
      child{ node[inner ysep=-1mm] (xil7) {$\ind{b}{5}$} }
      child{ node[inner ysep=-1mm] (xil8) {$\ind{c}{8}$} } } }
    child[level distance= 1cm]{ node (xil9) {$\ind{c}{7}$} };
\node[above left= of xiroot] {$\xi \in \CT:$};
\end{scope}

\begin{scope}[mygray]
\foreach[count=\i] \x  in {
    $\ind{a}{3}$,$\ind{b}{6}$,$\ind{a}{4}$,
    $\ind{a}{5}$,$\ind{b}{8}$,$\ \ \ind{c}{11}$,
    $\ind{b}{7}$,$\ \ \ind{c}{10}$,$\ind{c}{9}$}
  \node[below= 0.7mm of xil\i.center,fill=white,inner ysep=-.5mm, outer ysep=1.5mm, inner xsep=-0.5mm] (xi2l\i) {\x};
\node at (xi2root) {$d$};
\node[yshift=.5mm] at (xi2n1) {$e$};
\node at (xi2n2) {$d$};
\node at (xi2n21) {$d$};
\node[yshift=.5mm] at (xi2n22) {$e$};
\node[above right= of xiroot, xshift=1mm] (xi2name) {$\xi' \in \CT:$};
\end{scope}

\begin{scope}[xshift=10cm,
              level distance= 1.8cm,
              level 1/.style={sibling distance=40mm},
              level 2/.style={sibling distance=30mm},
              level 3/.style={sibling distance=16mm}]
\node (rhoroot) {$\bigl( q_f,(x^1_1 \, x^1_2 \, x^2_1 \, x^2_2 \, x^1_3 \, x^3_2) \bigr)$}
  child[level distance= 1.2cm]{ node {$\bigl(q_l,(x^1_1, \, x^1_2)\bigr)$} 
    child[sibling distance=16mm]{ node[xshift=-1mm] (rhol1) {$q_a$} }
    child[sibling distance=16mm]{ node[xshift=1mm] (rhol2) {$q_b$} } }
  child[xshift=0mm]{ node {$\bigl(q,(x^1_1 \, x^1_2, \, x^1_3 \, x^2_2, \, x^2_3 \,  x^3_2)\bigr)$} 
    child[xshift=6mm,level distance= 1.5cm]{ node (rhol3) {$q_a$} }
    child{ node {$\bigl(q,(x^1_1, \, x^1_2, \, x^1_3)\bigr)$} 
      child[level distance=1.4cm]{ node[xshift=-2mm] (rhol4) {$q_a$} }
      child[level distance=1.4cm]{ node (rhol5) {$q_b$} }
      child[level distance=1.4cm]{ node[xshift=2mm] (rhol6) {$q_c$} }  }
    child[xshift=3mm,level distance= 1.2cm]{ node {$\bigl(q_r,(x^1_1, \, x^1_2)\bigr)$}
      child{ node[xshift=-1mm] (rhol7) {$q_b$} }
      child{ node[xshift=1mm] (rhol8) {$q_c$} } } }
  child[xshift=-8mm,level distance= 1.2cm]{ node (rhol9) {$q_c$} };
\node[above left= of rhoroot] (rhoname) {$\rho \in \runs{\cA}:$};
\end{scope}

\begin{scope}[yshift=-8.5cm, xshift=10cm,
              level distance= 2cm,
              level 1/.style={sibling distance=50mm},
              level 2/.style={sibling distance=40mm},
              level 3/.style={sibling distance=16mm}]
\node (droot) {$q_f \to \bigl(d, (x^1_1 \, x^1_2 \, x^2_1 \, x^2_2 \, x^1_3 \, x^3_2 )\bigr)(q_l, q, q_c)$}
  child[level distance= 1.2cm]{ node {$q_l \to \bigl(e,(x^1_1, \, x^1_2)\bigr)(q_a, q_b)$} 
    child[sibling distance=16mm]{ node (dl1) {$q_a \to (a)$} }
    child[sibling distance=16mm]{ node (dl2) {$q_b \to (b)$} } }
  child[xshift=0mm]{ node {$q \to \bigl(d,(x^1_1 \, x^1_2, \, x^1_3 \, x^2_2, \, x^2_3 \,  x^3_2)\bigr)(q_a, q, q_r)$} 
    child[xshift=9mm,level distance= 1.2cm]{ node (dl3) {$q_a \to (a)$} }
    child{ node {$q \to \bigl(d,(x^1_1, \, x^1_2, \, x^1_3)\bigr)(q_a, q_b, q_c)$} 
      child[level distance=1.6cm]{ node (dl4) {$q_a \to (a)$} }
      child[level distance=1.6cm]{ node (dl5) {$q_b \to (b)$} }
      child[level distance=1.6cm]{ node (dl6) {$q_c \to (c)$} }  }
    child[xshift=1mm,level distance= 1.2cm]{ node {$q_r \to \bigl(e,(x^1_1, \, x^1_2)\bigr)(q_b, q_c)$}
      child[level distance=1.5cm]{ node (dl7) {$q_b \to (b)$} }
      child[level distance=1.5cm]{ node (dl8) {$q_c \to (c)$} } } }
  child[xshift=-5mm,level distance= 1.2cm]{ node (dl9) {$q_c \to (c)$} };
\node[above left= of droot] {$d \in \RT:$};
\end{scope}

\begin{scope}[yshift=-10.35cm,xshift=-1.25cm,
              execute at end node={\vphantom{$x^1_1$}},
              level distance= 1.5cm,
              level 1/.style={sibling distance=29mm},
              level 2/.style={sibling distance=20mm},
              level 3/.style={sibling distance=11mm}]
\node (troot) {$\bigl(d,(x^1_1 \, x^1_2 \, x^2_1 \, x^2_2 \, x^1_3 \, x^3_2 )\bigr)$}
  child[level distance= 1cm]{ node {$\bigl(e,(x^1_1, \, x^1_2)\bigr)$} 
    child[sibling distance=11mm]{ node[inner ysep=-0mm] (tl1) {$a$} }
    child[sibling distance=11mm]{ node[inner ysep=-0mm] (tl2) {$b$} } }
  child[xshift=0mm]{ node {$\bigl(d,(x^1_1 \, x^1_2, \, x^1_3 \, x^2_2, \, x^2_3 \,  x^3_2)\bigr)$} 
    child[xshift=4mm,level distance= 1cm]{ node (tl3) {$a$} }
    child{ node {$\bigl(d,(x^1_1, \, x^1_2, \, x^1_3)\bigr)$} 
      child[level distance=1.3cm]{ node (tl4) {$a$} }
      child[level distance=1.3cm]{ node (tl5) {$b$} }
      child[level distance=1.3cm]{ node (tl6) {$c$} }  }
    child[xshift=5mm,level distance= 1cm]{ node (t23) {$\bigl(e,(x^1_1, \, x^1_2)\bigr)$}
      child{ node[inner ysep=-0mm] (tl7) {$b$} }
      child{ node[inner ysep=-0mm] (tl8) {$c$} } } }
  child[level distance= 1cm]{ node (tl9) {$c$} };
\node[above left= of troot, xshift=2mm] {$t \in T_\Gamma:$};
\end{scope}

\node (aaa) at (-4,-7.8) {$(aaa bbb ccc)$};
\node (rep) at (0.5,-7.8) {rep$(\xi)$};

\node[draw, yscale=0.8,gray, inner sep=2mm, rounded corners=5mm, fit=(xi2name)(xil1)(rhol1)(rhol8)(rhol4)] (shape) {};
\node[fill=white,below left=5mm of shape.north east] {$\eqcl{\xi, \rho}$};

\begin{scope} [->,thick]
\draw (xi2l4.south west) ++(0,-0.2) to node[pos=0.7,right=3mm] {yield} ([yshift=4mm,xshift=4mm]aaa.north);
\draw (rep|-xi2l6.south) ++(0,-0.2) to node[pos=0.7,right] {rep} ([yshift=4mm]rep.north);

\draw (rep.west) ++(-0.35,0) to node[pos=0.45,above] {p-yield} ([xshift=3.5mm]aaa.east);

\draw (troot.north) ++(-1,0.5) to node[midway,right=4mm] {$\lcfrshom$} ([yshift=-3mm,xshift=4mm]aaa.south);
\draw (troot.north) ++(1,0.5) to node[midway,right=2mm] {$\ctahom$} ([yshift=-3mm]rep.south);

\draw (rhol5|-shape.south) ++(0,-0.3) to node[midway,right] {$\psi$} ([yshift=4mm]droot.north);

\draw ([xshift=3.3cm,yshift=-2mm]t23.north east) to node[midway,above] {$(.)_{\Gamma}$} ([xshift=1.1cm,yshift=-2mm]t23.north east);
\end{scope}

\end{tikzpicture}

%% file: fig-inductive.tex
\begin{tikzpicture}[remember picture, scale=0.8, node distance= -5mm and 0.5cm,
                    yscale=0.8, myyscale/.style={yscale=0.8},
                    every node/.style={transform shape,yscale=1/0.8},
                    execute at end node={\strut},
                    myfill/.style={fill=white,rounded corners=1mm,inner xsep=0.8mm,outer xsep=0.2mm, inner ysep=-0.1mm, outer ysep=2.1mm}] 

\begin{scope}[every node/.append style={myfill},
              level distance= 1.9cm,
              level 1/.style={sibling distance=29mm},
              level 2/.style={sibling distance=20mm},
              level 3/.style={sibling distance=11mm}]
\node (xiroot) {$d$}
  child[level distance= 1.8cm]{ node (xi1) {$e$} 
    child[level distance= 2.5cm,sibling distance=19mm,xshift=-15mm]{ node[inner ysep=-1mm] (xil1) {$\ind{a}{1}$\subnode{i1}{}} }
    child[level distance= 2.5cm,sibling distance=19mm,xshift=-15mm]{ node[inner ysep=-1mm] (xil2) {$\ind{b}{4}$\subnode{i4}{}} } }
  child[xshift=0mm,myfill]{ node (xi2) {$d$}
    child[level distance= 1.8cm,xshift=3mm]{ node (xil3) {$\ind{a}{2}$\subnode{i2}{}} }
    child{ node (xi22) {$d$} 
      child[sibling distance=19mm]{ node (xil4) {$\ind{a}{3}$\subnode{i3}{}} }
      child[sibling distance=19mm]{ node (xil5) {$\ind{b}{6}$\subnode{i6}{}} }
      child[sibling distance=19mm]{ node (xil6) {$\ind{c}{9}$\subnode{i9}{}} }  }
    child[xshift=15mm]{ node (xi2r) {$e$}
      child[sibling distance=20mm,xshift=15mm]{ node[inner ysep=-1mm] (xil7) {$\ind{b}{5}$\subnode{i5}{}} }
      child[sibling distance=20mm,xshift=15mm]{ node[inner ysep=-1mm] (xil8) {$\ind{c}{8}$\subnode{i8}{}} } } }
    child[level distance= 1cm,xshift=10mm]{ node (xil9) {$\ind{c}{7}$\subnode{i7}{}} };
\node[above left= of xiroot] {$\xi \in \CT:$};
\end{scope}

\newlength\rundist
\setlength\rundist{0.6em}
\newlength\rundistleaf
\setlength\rundistleaf{0em}

\node[right=\rundist of xiroot,yshift=2mm] {$(q_f, (x_1^1$\subnode{i1a2}{}$ x_2^1$\subnode{i23a0}{}$ x_1^2$\subnode{i4a2}{}$ x_2^2$\subnode{i56a0}{}$ x_3^1$\subnode{i7a1}{}$ x_2^3$\subnode{i89a0}{}$))$};
\node[right=\rundist of xi1,yshift=-1mm] {$(q_l, (x_1^1$\subnode{i1a1}{}$, x_2^1$\subnode{i4a1}{}$))$};
\node[right=\rundist of xi2] {$(q, (x_1^1$\subnode{i2a1}{}$ x_2^1$\subnode{i3a2}{}$, x_3^1$\subnode{i5a2}{}$ x_2^2$\subnode{i6a2}{}$, x_3^2$\subnode{i8a2}{}$ x_2^3$\subnode{i9a2}{}$))$};
\node[right=\rundist of xi22] {$(q, (x_1^1$\subnode{i3a1}{}$, x_2^1$\subnode{i6a1}{}$, x_3^1$\subnode{i9a1}{}$))$};
\node[right=0.3em of xi2r] {$(q_r, (x_1^1$\subnode{i5a1}{}$, x_2^1$\subnode{i8a1}{}$))$};
\node[right=\rundistleaf of xil1] {$(q_a)$};
\node[right=\rundistleaf of xil2] {$(q_b)$};
\node[right=\rundistleaf of xil3] {$(q_a)$};
\node[right=\rundistleaf of xil4] {$(q_a)$};
\node[right=\rundistleaf of xil5] {$(q_b)$};
\node[right=\rundistleaf of xil6] {$(q_c)$};
\node[right=\rundistleaf of xil7] {$(q_b)$};
\node[right=\rundistleaf of xil8] {$(q_c)$};
\node[right=\rundistleaf of xil9] {$(q_c)$};

\begin{scope}
    \draw[->,gray] ($(i1)-(0.4,-0.4)$) to[out=90,in=270] ($(i1a1)-(0.5,0.2)$);
    \draw[->,gray] ($(i2)-(0.4,-0.4)$) to[out=90,in=270] ($(i2a1)-(0.5,0.2)$);
    \draw[->,gray] ($(i3)-(0.4,-0.4)$) to[out=90,in=270] ($(i3a1)-(0.5,0.2)$);
    \draw[->,gray] ($(i4)-(0.4,-0.4)$) to[out=90,in=270] ($(i4a1)-(0.5,0.2)$);
    \draw[->,gray] ($(i5)-(0.4,-0.4)$) to[out=90,in=270] ($(i5a1)-(0.5,0.2)$);
    \draw[->,gray] ($(i6)-(0.4,-0.4)$) to[out=90,in=270] ($(i6a1)-(0.5,0.2)$);
    \draw[->,gray] ($(i7)-(0.4,-0.4)$) to[out=90,in=270] ($(i7a1)-(0.7,0.2)$);
    \draw[->,gray] ($(i8)-(0.4,-0.4)$) to[out=90,in=270] ($(i8a1)-(0.5,0.2)$);
    \draw[->,gray] ($(i9)-(0.4,-0.4)$) to[out=90,in=270] ($(i9a1)-(0.6,0.2)$);
    \draw[->,gray] ($(i1a1)-(0.4,-0.4)$) to[out=90,in=270] ($(i1a2)-(0.5,0.2)$);
    \draw[->,gray] ($(i3a1)-(0.4,-0.4)$) to[out=90,in=270] ($(i3a2)-(0.5,0.2)$);
    \draw[->,gray] ($(i4a1)-(0.4,-0.4)$) to[out=90,in=270] ($(i4a2)-(0.5,0.2)$);
    \draw[->,gray] ($(i5a1)-(0.5,-0.4)$) to[out=90,in=270] ($(i5a2)-(0.6,0.2)$);
    \draw[->,gray] ($(i6a1)-(0.4,-0.4)$) to[out=90,in=270] ($(i6a2)-(0.6,0.2)$);
    \draw[->,gray] ($(i8a1)-(0.6,-0.4)$) to[out=90,in=270] ($(i8a2)-(0.7,0.2)$);
    \draw[->,gray] ($(i9a1)-(0.7,-0.4)$) to[out=90,in=270] ($(i9a2)-(0.8,0.2)$);
    \draw[->,gray] ($(i2a1)-(0.4,-0.4)$) to[out=90,in=270] ($(i23a0)-(0.5,0.2)$);
    \draw[->,gray] ($(i3a2)-(0.4,-0.4)$) to[out=90,in=270] ($(i23a0)-(0.5,0.2)$);
    \draw[->,gray] ($(i5a2)-(0.6,-0.4)$) to[out=90,in=270] ($(i56a0)-(0.7,0.2)$);
    \draw[->,gray] ($(i6a2)-(0.6,-0.4)$) to[out=90,in=270] ($(i56a0)-(0.7,0.2)$);
    \draw[->,gray] ($(i8a2)-(0.7,-0.4)$) to[out=90,in=270] ($(i89a0)-(0.8,0.2)$);
    \draw[->,gray] ($(i9a2)-(0.7,-0.4)$) to[out=90,in=270] ($(i89a0)-(0.8,0.2)$);
\end{scope}

\begin{scope}[every node/.append style={myfill},
              level distance= 1.5cm,
              level 1/.style={sibling distance=24mm},
              level 2/.style={sibling distance=20mm},
              level 3/.style={sibling distance=11mm}]
\node[right=10cm of xiroot] (xiproot) {$d$}
  child[level distance= 1cm,xshift=5mm]{ node (xil9) {$\ind{a}{1}$} }
  child[xshift=0mm,myfill]{ node {$d$} 
    child[xshift=-0mm,level distance= 1cm]{ node {$e$}
      child{ node[inner ysep=-1mm] (xil7) {$\ind{a}{2}$} }
      child{ node[inner ysep=-1mm] (xil8) {$\ind{b}{5}$} } }
    child{ node {$d$} 
      child{ node (xil4) {$\ind{a}{3}$} }
      child{ node (xil5) {$\ind{b}{6}$} }
      child{ node (xil6) {$\ind{c}{9}$} }  }
    child[xshift=-5mm]{ node (xil3) {$\ind{c}{8}$} } }
  child[level distance= 0.9cm]{ node {$e$} 
    child[level distance= 0.9cm,sibling distance=11mm]{ node[inner ysep=-1mm] (xil1) {$\ind{b}{4}$} }
    child[level distance= 0.9cm,sibling distance=11mm]{ node[inner ysep=-1mm] (xil2) {$\ind{c}{7}$} } };
\node[above left= of xiproot] {$\xi' \in \CT:$};
\end{scope}

\end{tikzpicture}